\documentclass[aps,prb,twocolumn,showkeys,amsmath,amssymb,longbibliography,superscriptaddress]{revtex4-1}
\usepackage[bookmarks=false,linkcolor=blue,urlcolor=blue,colorlinks,citecolor=blue]{hyperref}
\usepackage{amsfonts}
\usepackage{wasysym}
\usepackage{amssymb,amsmath}
\usepackage{graphicx}
\usepackage{tabularx}
\usepackage{bm}
\usepackage{color}

\usepackage{ulem}
\usepackage{braket}
\usepackage{dcolumn}
\usepackage{verbatim}

\newcommand{\pd}{{\phantom\dag}}

\begin{document}

\title{Topology and localization of a periodically driven Kitaev model}

\author{I. C. Fulga}
\affiliation{Department of Condensed Matter Physics, Weizmann Institute of Science, Rehovot 76100, Israel}
\affiliation{IFW Dresden and W{\"u}rzburg-Dresden Cluster of Excellence ct.qmat, Helmholtzstr. 20, 01069 Dresden, Germany}

\author{M. Maksymenko}
\affiliation{Department of Condensed Matter Physics, Weizmann Institute of Science, Rehovot 76100, Israel}
\affiliation{SoftServe Inc., Austin, TX, USA}

\author{M. T. Rieder}
\affiliation{Department of Condensed Matter Physics, Weizmann Institute of Science, Rehovot 76100, Israel}

\author{N. H. Lindner}
\affiliation{Physics Department, Technion, 320003, Haifa, Israel}

\author{E. Berg}
\affiliation{Department of Condensed Matter Physics, Weizmann Institute of Science, Rehovot 76100, Israel}
\affiliation{Department of Physics, University of Chicago, Chicago, Illinois 60637, USA}

% \pacs{
%      }


\begin{abstract}
Periodically driven quantum many-body systems support anomalous topological phases of matter, which cannot be realized by static systems. In many cases, these anomalous phases can be many-body localized, which implies that they are stable and do not heat up as a result of the driving. What types of anomalous topological phenomena can be stabilized in driven systems, and in particular, can an anomalous phase exhibiting non-Abelian anyons be stabilized?   We address this question using an exactly solvable, stroboscopically driven 2D Kitaev spin model, in which anisotropic exchange couplings are boosted at consecutive time intervals. The model shows a rich phase diagram which contains anomalous topological phases. We characterize these phases using weak and strong scattering-matrix invariants defined for the fermionic degrees of freedom. Of particular importance is an anomalous phase whose zero flux sector exhibits fermionic bands with zero Chern numbers,  while a vortex binds a pair of Majorana modes, which as we show support non-Abelian braiding statistics. We further show that upon adding disorder, the zero flux sector of the model becomes localized. However, the model does not remain localized for a finite density of vortices. Hybridization of  Majorana modes bound to vortices form ``vortex bands'', which delocalize by either forming Chern bands or a thermal metal phase. We conclude that while the model cannot be many-body localized, it may still exhibit long thermalization times, owing to the necessity to create a finite density of vortices for delocalization to occur.
\end{abstract}

\maketitle

\section{Introduction}
\label{sec:intro}

Periodic driving provides a versatile tool for inducing and controlling topological phenomena in quantum many-body systems. In many cases, the driven systems realize Floquet versions of topological insulators and superconductors\cite{Yao2007,Oka2009, Inoue2010, Lindner2011,
Lindner2013, Gu2011, Kitagawa2011, Kundu2013, Delplace2013, Katan2013,  Lababidi2014, Iadecola2013, Cayssol2013, Goldman2014, Grushin2014, Kundu2014, Titum2015, Bukov2015} and artificial gauge fields.\cite{Goldman2014}
Interestingly, periodically driven systems open the possibility of going beyond these static analogues and accessing novel anomalous phases available only in Floquet systems.\cite{Kitagawa2010, Rudner2013, Jiang2011, Titum2016, Fulga2016, Asboth2014, Carpentier2015, Nathan2015, Khemani2016, Roy2016, Else2016, Keyserlingk2016, Potter2016, Roy2016, Keyserlingk2016a, Else2016a, Zhang2017, Nathan2017, Kundu2017} Experimental realizations of topological phenomena in Floquet systems have been demonstrated in the solid state,\cite{Wang2013} cold atomic setups,\cite{Jotzu2014} and photonics networks.\cite{Rechtsman2013, Mukherjee2017, Maczewsky2017}

Many studies of Floquet topological systems have focused on various aspects of the non-interacting quasi-energy spectrum, such as topological invariants and topologically protected edge modes. What, if anything, of these topological characteristics survives when inter-particle interactions are added? Generically, isolated Floquet many-body systems absorb energy indefinitely, reaching an ``infinite temperature'' steady state\cite{Grushin2014, dalessio2014} in which all the topological properties are lost, even though the energy absorption rate can be parametrically small in special limits.\cite{Abanin2015, Abanin2017} In the presence of quenched disorder, this fate can be evaded if the system is many-body localized, such that energy absorption from the driving field is suppressed.\cite{Ponte2015, Abanin2016} Floquet systems which are many-body localized were shown to support symmetry-protected topological phases.\cite{Potter2016, Roy2016, Else2016}Interestingly, in many cases the anomalous topological phases, which are unique to Floquet systems, can be many-body localized even in the absence of any symmetry.\cite{Titum2016,Nathan2017a}

Of particular interest is the possibility to stabilize \textit{topologically ordered states} which are unique to Floquet systems. What new kinds of topological orders and non-Abelian anyons exist in the presence of periodic driving? Can such states be stabilized against the effect of heating if quenched disorder is added? In this work, we address these questions by studying a class of topological states that can be realized by an exactly solvable, periodically driven spin model. Our approach is to start from Kitaev's honeycomb model\cite{Kitaev2006} and make the exchange couplings time-dependent (as in Refs.~\onlinecite{Sato2014, Thakurathi2014, Bhattacharya2016, Po2017}). The model can be solved by a mapping to free fermions coupled to a static $Z_2$ gauge field.\cite{Kitaev2006, Nussinov2015} Ref.~[\onlinecite{Po2017}]  pointed out that it supports a new form of anomalous topological order dubbed ``Floquet enriched topological order''. In this work, we focus on the broader phase structure of the model and the possibility to stabilize these phases by utilizing many-body localization.

In the clean (disorder-free) case, we find multiple phases, including phases of either ``strong'' or ``weak'' topological character (with chiral or non-chiral protected edge modes, respectively). We focus our attention on the ``anomalous'' topological phase whose fermionic spectrum is characterized by the presence of chiral edge states, despite the fact that all the fermionic Floquet bulk bands are topologically trivial. The possibility of such a Floquet band structure has been discussed in Refs.~\onlinecite{Kitagawa2010, Rudner2013, Titum2016}. In this phase, the $Z_2$ flux excitations carry \textit{pairs} of localized Majorana modes at quasi-energies $0$ and $\pi/T$, where $T$ is the driving period, protected by the time periodicity of the fermionic Hamiltonian.\cite{Yang2018} We show that upon exchange, the two Majorana modes behave as two independent non-Abelian Ising anyons.

To investigate whether these phases can be stabilized using many-body localization, we study whether the full spectrum of the exactly solvable model can be localized in the presence of disorder. If that is indeed the case, then many-body localization may persist in the presence of integrability breaking perturbations. To this end, we add quenched disorder in a way that preserves the solvability of the model, and explore the localization properties of the fermionic spectrum in the presence of a static flux configuration.

Since in the anomalous phase the fermionic bulk Floquet bands have zero Chern numbers, \textit{all} the fermionic bulk states can be localized in the presence of quenched disorder.\cite{Titum2016} We demonstrate that disorder can be introduced into the driven spin model such that all states are localized if no $Z_2$ fluxes are present. In sectors with a finite density of fluxes, the Majorana modes bound to these fluxes form bands centered around quasi-energy $0$ and $\pi/T$. We find that these bands become delocalized, either forming a ``thermal metal'' phase~\cite{Laumann2012, Lahtinen2014} or by forming bands with non-zero Chern numbers. Hence, while in the flux-free sector of the theory all the fermionic excitations can be localized, the entire spectrum (including all flux sectors) necessarily contains delocalized ``single-particle'' states. This suggests that, in the presence of perturbations breaking the integrability of the driven Kitaev model, the anomalous topological phase cannot be fully localized. However, starting from a flux state with vanishing flux density,  as long as the rate for creating flux excitations is small, a system initialized in the flux-free sector may take a long time to thermalize.

The rest of the paper is organized as follows. In Section \ref{sec:flkit} we review the Kitaev honeycomb model, its mapping to free Majorana modes, and introduce the driving protocol. We study the phase diagram of the system (Section \ref{sec:pd}), which is periodic in driving strength, and discuss the presence of Floquet-Majorana modes bound to vortex defects (Section \ref{sec:majorana}). In Section \ref{sec:dis_numerics} we discuss the effect of disorder, focusing on particular points in the phase diagram at which all the states are localized, while in Section \ref{sec:vortexdens} we argue that delocalization necessarily occurs away from these points, given a finite density of flux excitations.
We conclude and discuss directions for future work in Section \ref{sec:conc}.

\section{Periodically driven Kitaev model}
\label{sec:flkit}

\begin{figure}[tb]
\begin{center}
\includegraphics[width=0.45\textwidth]{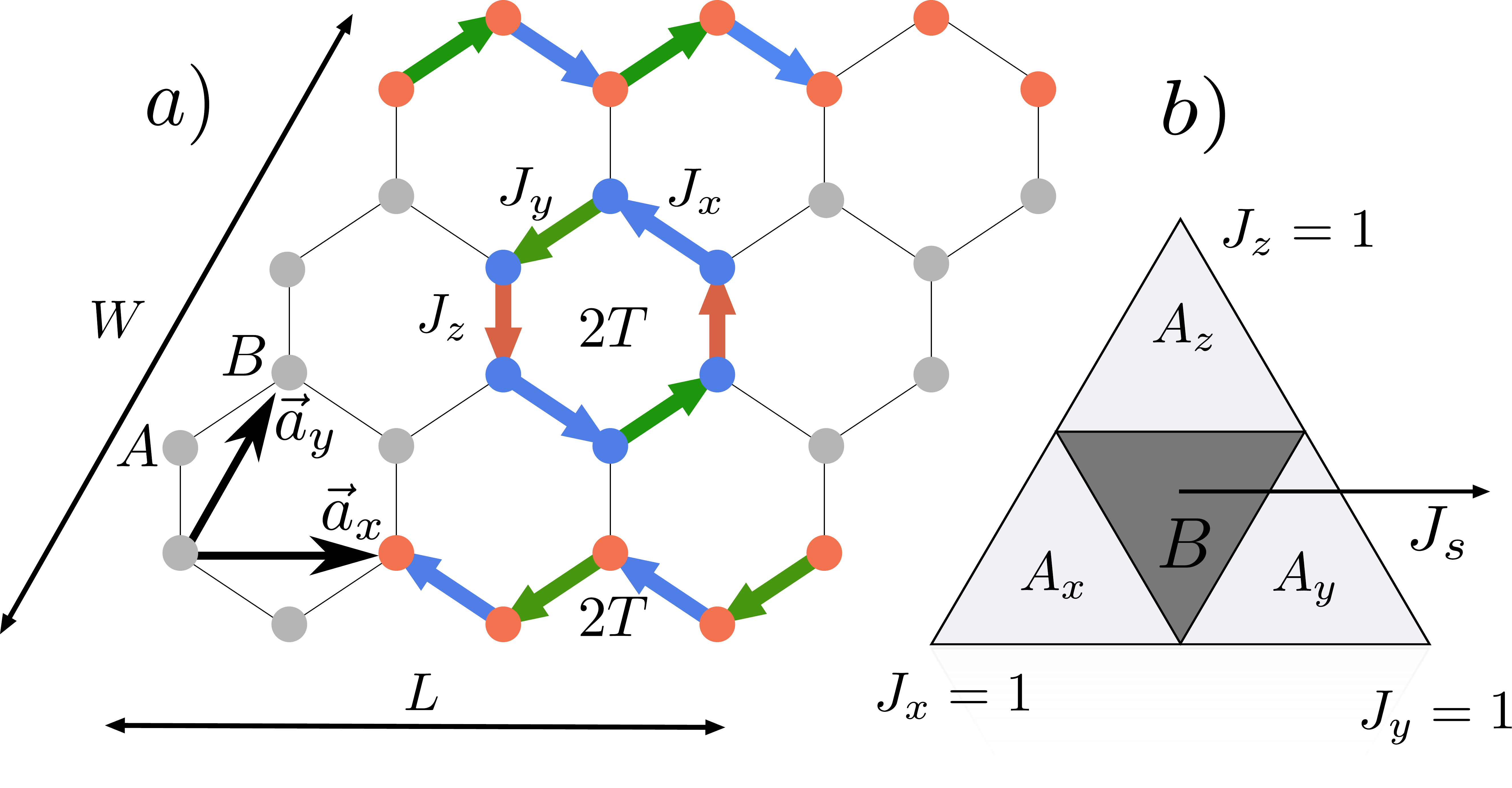}
\caption{a) Driving protocol of the 2D Kitaev model. We use a 2D $L \times W$  honeycomb lattice with Bravais vectors $\vec{a}_{x}$ and $\vec{a}_{y}$ and two sublattices labeled as $A$ and $B$. The driving protocol involves consecutive boosts of $J_{x},\;J_{y}$ and $J_{z}$. When $J_{s}T/4=\pi/2$ and $J_u=0$ Majorana operators return to their initial positions after two driving periods, $2T$. In a finite system, this will lead to the formation of chiral edge states at the boundaries. b) The phase diagram of the static Kitaev model consists of three gapped $A$-phases and a gapless $B$-phase. The driving strength $J_{s}$ acts as additional axis in this diagram.}
\label{fig:lattice_protocol}
\end{center}
\end{figure}

Kitaev's honeycomb model describes $S=1/2$ spins on a hexagonal lattice that are coupled in a highly anisotropic fashion.\cite{Kitaev2006} Here we consider a time-dependent version of the model, given by
\begin{equation}
H=\sum_{\langle j,k \rangle_\alpha}
J_{\alpha, jk} (t) \sigma^{\alpha}_{j}\sigma^{\alpha}_{k} \, ,
\label{kitaev_ham}
\end{equation}
where $\langle j,k \rangle_\alpha$ denotes a pair of neighboring spins on the honeycomb lattice which are connected by a bond of type $\alpha=x,y,z$ (see Fig.~\ref{fig:lattice_protocol}a). The Pauli matrices $\sigma^\alpha$ act on the spin degree of freedom, and $J_{\alpha, jk}(t)$ is the coupling strength.
The model Eq.~\eqref{kitaev_ham} describes a strongly interacting system, whose analysis, however, is simplified by the observation that there exists an integral of motion for each unit cell in the lattice.
For every hexagonal plaquette $p$, a so-called plaquette operator,
$\hat{W}_{p}$,
commutes with the Hamiltonian.\cite{Kitaev2006}
These mutually commuting operators $\hat{W}_{p}$ have eigenvalues $w_{p}=\pm 1$ and the total Hilbert space of the Hamiltonian Eq.~\eqref{kitaev_ham} decomposes into sectors that are distinguished by those eigenvalues, in the following referred to as \textit{flux sectors}.

The Hamiltonian Eq.~(\ref{kitaev_ham}) can be mapped into a problem of Majorana modes coupled to a static $Z_2$ gauge field as follows.\cite{Kitaev2006} For each spin of the lattice, located on site $j$, one introduces four Majorana operators, $c_j$ and $b_j^{\alpha_{jk}}$, with $\alpha_{jk}=x,y,z$ depending on the orientation of the bond between neighboring spins $j$ and $k$. Then, by identifying $\sigma_j^\alpha = i b_j^{\alpha_{jk}} c_j$, the fermionic Hamiltonian can be written as
\begin{equation}
H_{u}=\frac{i}{2} \sum_{\langle j,k \rangle_\alpha} J_{\alpha, jk}(t) u_{jk}c_{j}c_{k},
\label{eq:ham_majorana}
\end{equation}
where $u_{jk}=ib^{\alpha_{jk}}_jb^{\alpha_{jk}}_k$ is the $Z_2$ gauge field defined on the bond connecting the sites $j$ and $k$. The  $Z_2$ flux $w_p$ through a given plaquette is equal to the product of the $u_{jk}$'s encircling that plaquette.\footnote{The Hilbert space that the $c_j$ and $u_{jk}$ act on is larger than the Hilbert space of the spins, thus a projection on the physical Hilbert space is required, see Ref.~\onlinecite{Kitaev2006}}
Since the gauge field operators $u_{jk}$ commute with $H$, we can replace them with their eigenvalues, $u_{jk}=\pm 1$. This is equivalent to working in a given flux sector, in a fixed gauge.

For time-independent $J_{\alpha,jk}$ the phase diagram of the model is shown in Fig.~\ref{fig:lattice_protocol}b: in the so-called $A$-phase, $|J_{x}|+|J_{y}|<|J_{z}|$, the system has a gapped spectrum and exhibits a $Z_{2}$ topological order with Abelian anyonic excitations. In contrast, in the $B$-phase, $|J_{x}|+|J_{y}|>|J_{z}|$, the system is gapless, but can be brought into a topological state with non-Abelian statistics upon opening a gap with a time-reversal symmetry-breaking perturbation.\cite{Kitaev2006}

In this paper, we explore the case where the $J_{x,y,z}$ depend periodically on time and investigate the various emerging phases of the model.\cite{Sato2014, Po2017}
We consider a driving protocol in which each coupling  can be written as $J_\alpha=J_u+J_s(t)$, where $ J_u $ is time-independent and $J_s(t)$ is piecewise constant. Specifically, we use a four-step driving protocol\cite{Kitagawa2010}

\begin{flalign}\label{eq:driving_protocol}
J_x = & J_s + J_u, J_{y,z}=J_u \\
& \text{ for } nT < t \leq nT+ T/4, \nonumber\\
  \nonumber\\
J_y = & J_s + J_u, J_{z,x}=J_u \\
& \text{ for } nT+T/4< t \leq nT+ T/2,\nonumber\\
  \nonumber\\
J_z = & J_s + J_u, J_{x,y}=J_u \\
& \text{ for } nT+T/2< t \leq nT+ 3T/4, \nonumber\\
  \nonumber\\
J_{x,y,z} = & J_u \\
& \text{ for } nT+3T/4< t \leq nT+ T, \nonumber
\end{flalign}
% \begin{enumerate}\label{eq:driving_protocol}
% \item $J_x = J_s + J_u$, $J_{y,z}=J_u$
% 
% for $nT< t \leq nT+ T/4$,\\
% \item $J_y = J_s + J_u$, $J_{z,x}=J_u$
% 
% for $nT+T/4< t \leq nT+ T/2$,\\
% \item $J_z = J_s + J_u$, $J_{x,y}=J_u$
% 
% for $nT+T/2< t \leq nT+ 3T/4$,
% \item $J_{x,y,z} = J_u$
% 
% for $nT+3T/4< t \leq nT+ T$,
% \end{enumerate}
with $T$ the driving period, and $n\in\mathbb{Z}$. The Hamiltonian Eq.~\eqref{eq:ham_majorana} is now time-periodic, $H_u(t) = H_u(t+T)$, such that its behavior can be characterized by studying the time-evolution operator over one driving period, known as the Floquet operator. Setting $\hbar=1$, the latter reads
\begin{equation}\label{eq:flop}
F = {\cal T} \exp \left( -i \int_0^{T} H_u(t)dt \right),
\end{equation}
where ${\cal T}$ denotes time-ordering. The eigenvalues and eigenvectors of $F$ play a similar role to the energies and wavefunctions of static systems, with one important difference. Owing to the periodic nature of the driving field, eigenvalues of the unitary Floquet operator take the form $\exp (-i\varepsilon T)$, with $\varepsilon$ referred to as quasi-energy. Unlike energy levels in time-independent systems, the quasi-energy is periodic, $\varepsilon = \varepsilon + 2\pi/T$.

The specific form of the driving protocol simplifies the evaluation of the Floquet operator Eq.~\eqref{eq:flop}, which in our case takes the form
\begin{align}\label{eq:floquet_decompose}
 F &= F_{4} \,  F_{3} \,  F_{2} \,  F_{1}, \\
 F_{i} &=   \exp( -i H_i T/4),
\end{align}
where $H_i$ are the Hamiltonians for the four steps of the driving protocol shown above.

\section{Phase diagram}
\label{sec:pd}

\begin{figure*}[t]
\includegraphics[width=0.8\textwidth]{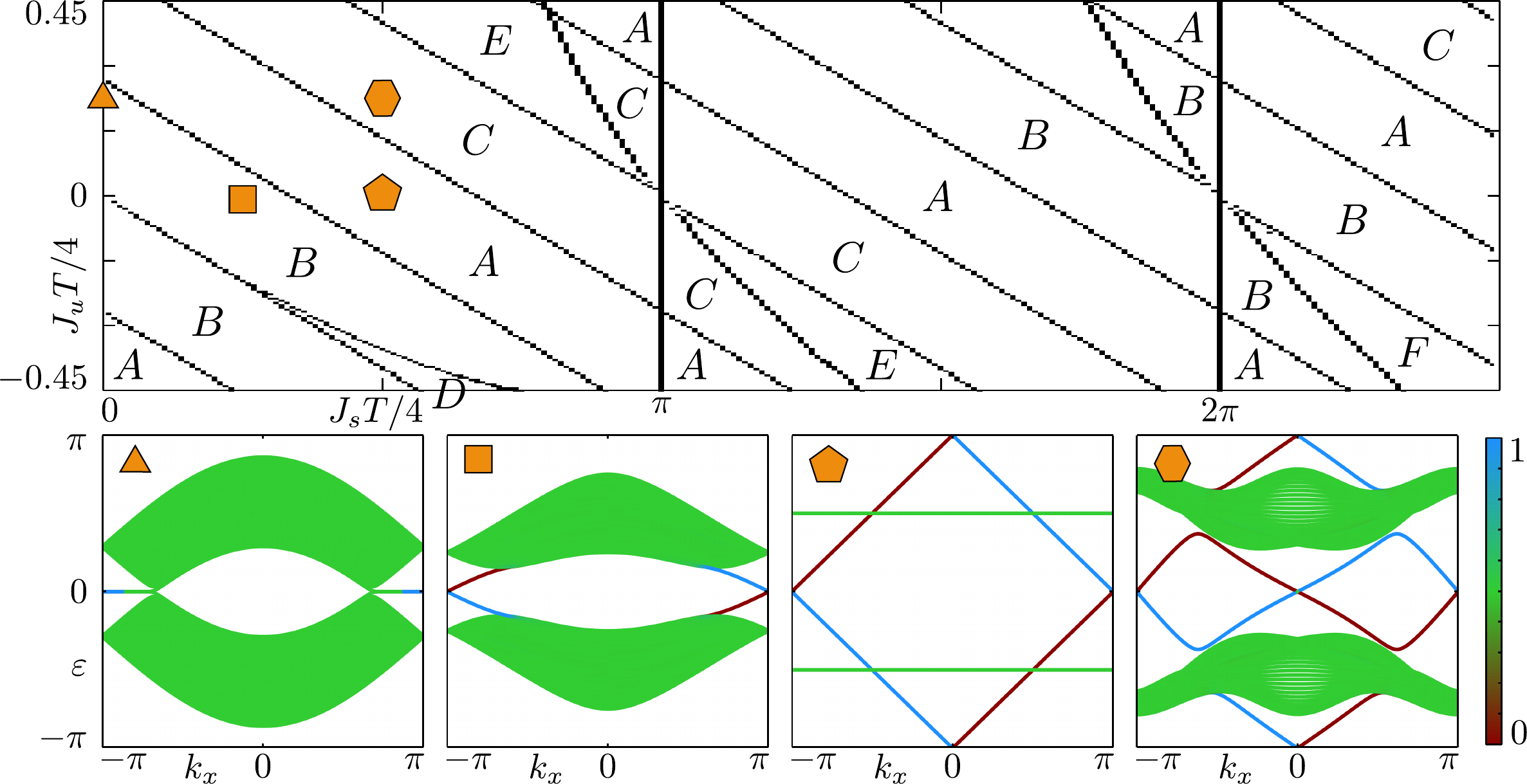}
\caption{Top:
phase diagram mapped by determining gap closings at quasi-energies $\varepsilon=0$ and $\varepsilon=\pi$. Each phase is labeled by its strong and weak topological invariants at $\varepsilon=0,\pi$, $(\mathcal{W}_0, \nu_{x,0}, \mathcal{W}_\pi, \nu_{x,\pi})$, computed using the scattering matrix formalism (see Appendix \ref{app:scattering_matrix}). The invariants are: $(-1,-1,-1,1)$ in phase $A$, $(-1,-1,0,1)$ in phase $B$, $(0,-1,-1,1)$ in phase $C$, $(1,-1,0,1)$ in phase $D$, $(0,-1,3,1)$ in phase $E$, and $(3,-1,0,1)$ in phase $F$.
The pattern of topological phases repeats when increasing $J_s$ due to the periodicity of the system in driving strength, Eq.~\eqref{eq:flper}.
Bottom: bandstructures of the system in an infinite ribbon geometry (infinite along $\vec{a}_x$, with $W=40$). The color scale denotes the wavefunction amplitude on the bottom 20 lattice sites, meaning that bulk states are plotted in green, whereas modes on the top and bottom boundaries are plotted in red and blue, respectively. From left to right, the panels show the bandstructure computed for values of $J_s$ and $J_u$ marked by orange symbols in the top panel: $(J_uT/4, J_sT/4)=(0.225, 0)$ $\triangle$, $(0,\pi/4)$ $\Square$, $(0,\pi/2)$ $\pentagon$, and $(0.225,\pi/2)$ $\hexagon$.}
\label{fig:phase_diag}
\end{figure*}

The Floquet operator Eq.~\eqref{eq:flop} breaks time-reversal symmetry due to the structure of the driving protocol. However, since the Hamiltonian Eq.~\eqref{eq:ham_majorana} describes a lattice model of Majorana modes, the fermionic spectrum shows a particle-hole symmetry taking the real-space form
\begin{equation}\label{eq:ph_sym_floquet_realspace}
F=F^*,
\end{equation}
which is valid for any configuration of the static $Z_2$ gauge field and therefore applies to any vortex sector.

In the vortex free sector, translational symmetry enables writing the Hamiltonian in momentum space. Using the basis of Fig.~\ref{fig:lattice_protocol}, the Hamiltonian reads
\begin{equation}\label{eq:ph_ham_big}
H_u =\sum_{{\bf{k}}}
\begin{pmatrix}
 c_{A,{-\bf{k}}} & c_{B,{-\bf{k}}}
\end{pmatrix}
{\cal H}_{\bf k}
\begin{pmatrix}
c_{A,{\bf{k}}} \\
c_{B,{\bf{k}}}
\end{pmatrix},
\end{equation}
with
\begin{equation}\label{eq:ph_ham}
\begin{split}
{\cal H}_{\bf k} =& (J_x + J_y \cos(k_x) + J_z \cos(k_y) ) \tau_y \\
 & + (J_y \sin(k_x) + J_z \sin(k_y)) \tau_x,
\end{split}
\end{equation}
and
\begin{equation}
 c_{A/B,{\bf k}}=\frac{1}{\sqrt{2N}}\sum_{\bf r} e^{i {\bf k}{\bf r}}c_{A/B,{\bf r}},
\end{equation}
where ${\bf r}$ denotes the position of the unit cell, $A$ and $B$ are the sub-lattices, $N$ is the total number of unit cells, and the Pauli matrices $\tau_i$ act on the sublattice degree of freedom. In momentum space, particle-hole symmetry takes the form 
\begin{equation}\label{eq:ph_floquet_momentum_space}
 F({\bf k})=F^*(-{\bf k}).
\end{equation}

We begin by considering the vortex-free sector and later study the effect of a nonzero number of $Z_2$ flux excitations. While in static systems the particle-hole symmetry relates states with positive and negative energies, in driven systems there are two particle-hole symmetric quasi-energies, $\varepsilon=0,\pi$, left invariant by the transformation Eq.~\eqref{eq:ph_floquet_momentum_space}, which maps $\varepsilon({\bf k})$ to $-\varepsilon(-{\bf k})$.

For generic values of the parameters $J_u$ and $J_s$ the quasi-energy spectrum of a system with periodic boundary conditions consists of two quasi-energy bands separated by gaps at $\varepsilon=0$ and $\varepsilon=\pi$. As a function of the coupling parameters $J_u$ and $J_s$, the system undergoes a series of topological phase transitions, marked by the closing and reopening of these quasi-energy gaps.

We determine the phase boundaries by locating the lines of gap-closing in the $(J_u,J_s)$ plane. The resulting phase diagram is shown in Fig.~\ref{fig:phase_diag}a. To determine the nature of each phase, we use the Kwant code\cite{Groth2014} to study the system in the infinite ribbon geometry, (infinite along $\vec{a}_x$ and finite along $\vec{a}_y$, see Fig.~\ref{fig:lattice_protocol}). Each phase is characterized by the strong topological invariants $\mathcal{W}_0$ and $\mathcal{W}_\pi$, which
count the net number of chiral edge states (on each edge of the ribbon) within the gap at $\varepsilon=0$ and $\varepsilon=\pi$, respectively. These topological invariants may be determined either from the full time evolution operator of the bulk system, $U({\bf k}, t)$,\cite{Rudner2013} or from the scattering matrix of a Floquet system coupled to leads, as done in Appendix \ref{app:scattering_matrix}. The strong topological invariants are related to the Chern numbers of the two quasi-energy bands as 
\begin{equation}
C_\pm=\pm\left(\mathcal{W}_\pi-\mathcal{W}_0\right)
\end{equation}
where $C_\pm$ are the Chern numbers of the Floquet bands $|u_+\rangle$ and $|u_-\rangle$, with quasi-energy $-\pi<\varepsilon<0$ and $0<\varepsilon<\pi$:
\begin{equation}
 C_\pm = \frac{i}{2\pi} \int d^2k \left( \left\langle \frac{\partial u_\pm}{\partial k_x} \middle| \frac{\partial u_\pm}{\partial k_y} \right\rangle - \left\langle \frac{\partial u_\pm}{\partial k_y} \middle| \frac{\partial u_\pm}{\partial k_x} \right\rangle \right).
\end{equation}

In addition to the strong invariants $\mathcal{W}_{0,\pi}$, each phase is also characterized by four weak invariants, $\nu_{j, \varepsilon}$, with $j=x,y$ and $\varepsilon=0,\pi$. The latter require both particle-hole symmetry and translation symmetry, and determine the positions of edge modes in the Brillouin zone in the same way as for time-independent topological superconductors.\cite{Asahi2012, Seroussi2014} In a ribbon geometry with edges parallel to $a_j$ ($j=x,y$), the index $\nu_{j,\varepsilon}$ counts the parity of the number of edge modes crossing $k_j=\pi$ at each of the two particle-hole symmetric quasi-energies, $\varepsilon=0,\pi$. For instance, in a ribbon geometry with edges parallel to $a_x$, a system with $\mathcal{W}_\pi=1$ will have one protected chiral edge mode crossing the quasi-energy zone edge, $\varepsilon=\pi$. Due to the constraint imposed on the spectrum by particle-hole symmetry, the edge mode can cross either at $k_x=0$, implying that $\nu_{x,\pi}=1$, or at $k_x=\pi$, in which case $\nu_{x,\pi}=-1$. The weak invariants in different directions are always equal, $\nu_{x,\varepsilon}=\nu_{y,\varepsilon}$ (see Appendix \ref{app:ribbon}),  such that each phase of the model can be uniquely identified by four invariants, $(\mathcal{W}_0, \nu_{x, 0}, \mathcal{W}_\pi, \nu_{x,\pi})$.

We now describe the phase diagram in more detail.
For $J_{s}=0$ (Fig.~\ref{fig:phase_diag}, marked by $\triangle$), the system is time-independent. In this case, we recover the graphene-like spectrum of the static Kitaev model in the gapless phase.\cite{Kitaev2006}

Including a driving component $J_{s}\neq0$ leads to a gap opening and the quasi-energy bands acquire a non-zero Chern number (Fig.~\ref{fig:phase_diag}, phase $B$).

For $J_{u}T/4=0$ and $J_{s} T/4 = \pi/2$ (Fig.~\ref{fig:phase_diag}, marked by $\pentagon$), during the first three steps of the driving protocol the system consists of Majorana dimers which are decoupled from each other. In every one of these three steps, the two connected Majoranas of a given dimer, $c_1$ and $c_2$, are swapped, transforming the fermionic state as $c_1\to c_2$ and $c_2\to -c_1$, which can be verified by direct calculation. Therefore, after two full driving periods each operator comes back to itself, as shown in Fig.~\ref{fig:lattice_protocol}. We show in more detail in the next section that this implies the Floquet-single particle states are localized, forming flat bands which have zero Chern number and are positioned at quasienergies $\varepsilon=\pm\pi/2$. For a finite system however, states at the edge of the system move by one unit cell every period, forming dispersing edge modes in both $\varepsilon=0$ and $\varepsilon=\pi$ quasi-energy gaps. The appearance of chiral edge states in this case is unique to driven systems,\cite{Rudner2013} as in the static case they cannot occur when all the bulk bands have zero Chern numbers.\cite{Bergholtz2013, Derzhko2015, Lee2016} We will refer to this phase the ``anomalous Floquet topological phase'' (or just the ``anomalous phase'').

At the point $J_{u}T/4= 0.225$ and $J_{s}T/4=\pi/2$ (Fig.~\ref{fig:phase_diag}, marked by $\hexagon$) the system is in a weak topological phase at $\varepsilon=0$ and a strong phase at $\epsilon=\pi$. The weak phase exhibits counter-propagating Majorana edge modes protected by a combination of particle-hole symmetry and translation symmetry in either the $\vec{a}_x$ or the $\vec{a}_y$ directions. Since the time-independent coupling $J_u$ and the driving strength $J_s$ are isotropic, counter-propagating Majorana edge modes appear both on boundaries parallel to $\vec{a}_x$ or $\vec{a}_y$ (see Appendix \ref{app:ribbon}).

As $J_s$ is increased, the phase diagram of Fig.~\ref{fig:phase_diag} shows repeating patterns. This is due to the periodicity of the model in driving strength. In the absence of a time-independent coupling $J_u=0$, the Floquet operator obeys
\begin{equation}\label{eq:flper}
 { F}(J_s T/4) = -{ F}(J_s T/4+\pi)={ F}(J_s T/4+2\pi),
\end{equation}
such that the topological phases of the model repeat in a periodic fashion. This enables realizing strong and weak topological phases for arbitrarily slow driving (See Appendix \ref{apdx:periodic_phase_diag}).

\section{Majorana modes at vortex defects}
\label{sec:majorana}

\begin{figure}[tb]
\begin{center}
\includegraphics[width=0.45\textwidth]{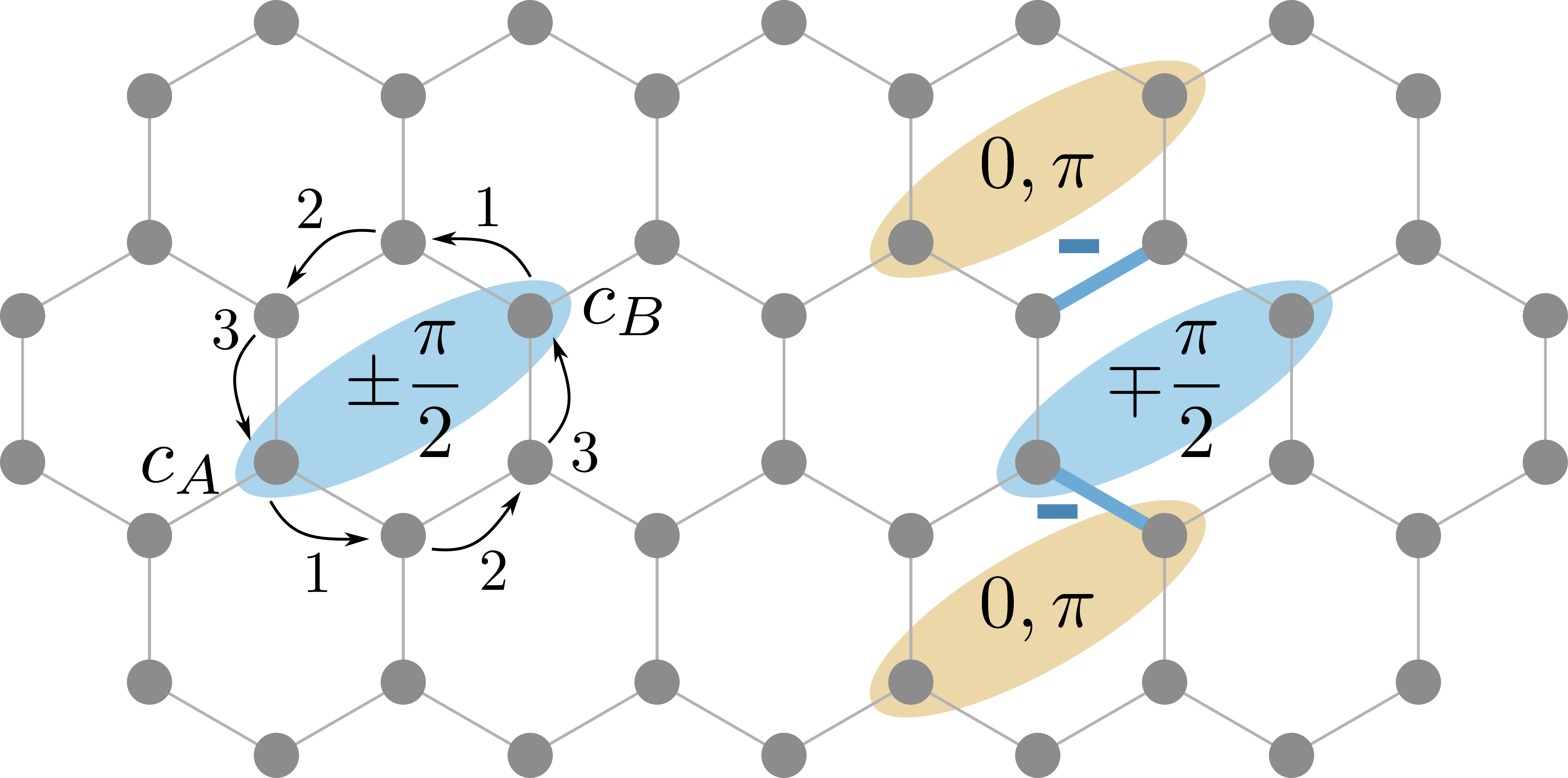}
\caption{Dynamics and quasi-energy spectrum in the anomalous phase with $J_u=0$, $J_s=\pi/2$. On the left, we show the evolution of two Majorana operators $c_A$, $c_B$. During the first three parts of the period, labeled $1$ to $3$, the Majorana operators are transferred to a neighboring site, as shown by the arrows. After a full period of the time evolution, $c_A$, $c_B$ are swapped.
On the right, a pair of vortices is introduced by flipping the sign of two bonds (marked with thick lines). States on plaquettes with an even number of flipped bonds (blue ellipses) have energies $\varepsilon=\pm\pi/2$. Each vortex is modeled as a plaquette with an odd number of flipped bonds (yellow ellipses), and binds two Floquet-Majorana states, at quasi-energies $\varepsilon=0$ and $\varepsilon=\pi$.}
\label{fig:kitaev_vortex}
\end{center}
\end{figure}

In the static Kitaev model, a phase with a non-zero Chern number can occur when time-reversal symmetry is broken. Then, vortex defects in the $Z_{2}$ gauge field $\hat{u}_{ij}$ bind Majorana states at $\varepsilon=0$.\cite{Kitaev2006} In contrast, in the driven Kitaev model there are two particle-hole symmetric quasi-energies and therefore
isolated Majorana modes can appear either at quasi-energy $\varepsilon=0$ or $\pi$.\cite{Thakurathi2013} As we show below, this happens at the vortex cores in the anomalous phase $(W_0=W_\pi \ne 0)$.

The fact that vortices in the anomalous phase carry Majoranas at both $0$ and $\pi$ quasi-energies can be seen analytically, considering the special case of resonant stroboscopic driving, $J_{s} T/4 = \pi/2$, and a vanishing uniform component $J_{u}=0$. At this point the quasi-energy spectrum of the driven Kitaev model can be analyzed exactly. In each of the first three steps of the driving protocol, the Majorana operators connected by a non-zero hopping are swapped. Given the order of steps in the driving protocol, a full period of the time evolution effectively swaps the two operators $c_A$ and $c_B$ located on opposite sites of a given hexagonal plaquette, as shown in Fig. \ref{fig:kitaev_vortex}. The Floquet eigenmodes are therefore superpositions $(c_A \pm  i c_B)/\sqrt{2}$ of wavefunctions located at the two sites. Upon applying one period of the time evolution, those states acquire phase-factors
\begin{align}
\frac{c_A \pm  i c_B}{\sqrt{2}} \to \pm i \frac{c_A \pm  i c_B}{\sqrt{2}}
\label{eq:pm}
\end{align}
corresponding to quasi-energies $\pm \pi/2$, as can be verified by direct calculation.

Introducing vortices (i.e. $Z_2$ fluxes) into the system amounts to locally flipping signs in the coupling terms $J_{x,y,z}$. A hexagonal plaquette is said to carry a $\pi$-vortex if the couplings have switched sign at an odd number of its bonds. Thus, vortices appear in pairs but can proliferate through the system while being connected by a string of flipped couplings (see Fig. \ref{fig:kitaev_vortex}). We focus on the resonant driving case, $J_sT/4 = \pi/2$, and analyze the quasi-energy spectrum in the presence of two types of plaquettes: one with an even number of couplings of the same sign, and one with an odd number of flipped bonds. Let $n_A$ and $n_B$ be the number of flipped bonds encountered by the operators $c_A$ and $c_B$ during the first three steps of the driving protocol (see Fig.~\ref{fig:kitaev_vortex}). By explicitly computing the Floquet operator of a single plaquette, we find that the Majorana operators transform as
\begin{equation}\label{eq:pm_vortex}
 \frac{c_A \pm  i c_B}{\sqrt{2}} \to \pm i \frac{(-1)^{n_B}c_A \pm  i (-1)^{n_A}c_B}{\sqrt{2}}
\end{equation}
during a full driving cycle. For a plaquette containing an even number of flipped bonds, Eq.~\eqref{eq:pm_vortex} is identical to the evolution obtained without any flipped bonds [Eq.~\eqref{eq:pm}] if both $n_A$ and $n_B$ are even, or leads to a global minus sign if $n_A$ and $n_B$ are both odd. In both cases, the combination $(c_A \pm  i c_B)/\sqrt{2}$ is still a fermionic eigenmode with quasienergy $\pi/2$ or $-\pi/2$, depending on the parity of $n_{A,B}$. A different situation is realized in a plaquette with an odd number of flipped bonds. In that case, the complex superposition of the two operators $c_A$ and $c_B$ is no longer an eigenmode of the Floquet operator. We find instead that the eigenmodes are formed by the real superposition $(c_A\pm c_B)/\sqrt{2}$, and the corresponding quasienergies are $\pi, 0$. The resulting vortex states are Majoranas, which cannot shift away from $\varepsilon = 0, \pi$ due to particle-hole symmetry, and must therefore remain localized as long as the system is in the anomalous phase. In Appendix \ref{app:braiding} we discuss the braiding properties of the two types of Majorana modes, showing that they behave as two independent sets of anyons.

\section{Effects of disorder}
\label{sec:dis}

\subsection{Model and numerical study}
\label{sec:dis_numerics}

Floquet many-body systems generically absorb energy indefinitely and all topological features are lost in the steady state,\cite{Grushin2014, dalessio2014} unless the system is many-body localized.\cite{Ponte2015, Abanin2016} Certain anomalous chiral Floquet phases have been shown to be compatible with many-body localization .\cite{Po2016, Harper2017, Nathan2017}
Can the Floquet-Kitaev model be many-body localized in the presence of disorder? To answer this question, we first examine whether the fermionic spectrum of the exactly-solvable Floquet-Kitaev model is fully localized, and at the same time, the system is in an anomalous phase. If this is indeed possible, then many-body localization may persist in the presence of interactions that spoil the exact solvability of the model.

Quasi-energy bands which have a non-zero Chern number cannot be spanned by a complete basis of localized Wannier functions,\cite{Thouless1984, Thonhauser2006} implying the existence of at least one mobility edge as a function of quasi-energy.\cite{Halperin1982} As such, any phase in which bands carry non-zero Chern numbers necessarily contains delocalized states, independently of the type of disorder. In contrast, for the anomalous topological phase with trivial bulk bands, it was shown that disorder can lead to an entirely localized bulk coexisting with chiral edge states, which are completely decoupled from the bulk at all quasi-energies.\cite{Titum2016} The phase was dubbed an \textit{anomalous Floquet Anderson insulator} (AFAI), and is only accessible by periodic driving.

Of all the phases appearing in the driven Kitaev model (see Fig~\ref{fig:phase_diag}), thus, only the anomalous phase (phase $C$) could possibly be stable against heating and will therefore be studied hereafter. While the free-fermion problem in the flux-free sector is similar to that of Ref.~\onlinecite{Titum2016}, localization has to occur in every vortex sector in order to guarantee that the system does not absorb energy in a generic (not fine-tuned) Hamiltonian.

To determine the robustness of the Floquet-Kitaev model in the anomalous phase, we study the localization properties of the free fermion system Eq.~\eqref{eq:ham_majorana} in the presence of disorder which maintains the exact solvability of the model. If the latter leads to localized bulk states at any quasi-energy and density of vortex excitations, we
expect the disordered phase of the periodically driven Kitaev model to be stable against heating, allowing its topological properties to be observed.

We start by considering the system at the point of resonant driving, i.e. $J_u =0$ and $J_s T/4 =\pi/2$, where the Floquet bulk bands are completely flat since the corresponding states are strictly localized on opposite sites of each hexagonal plaquette.
When introducing vortices into the system Majorana modes are formed, which, like the bulk states, are strictly localized on the vortex plaquettes (see Fig.~\ref{fig:kitaev_vortex}). At finite vortex densities, they form flat bands at quasi-energies $\varepsilon= 0, \pi$. For small deviations away from the resonant driving point, the bulk bands acquire a dispersion. Similarly, Majorana modes overlap and form dispersing bands, whose form may depend strongly on the configuration of the vortices. In the following we will consider random configurations of fluxes arising from disorder in the system, showing that away from the resonant driving point the system will necessarily host delocalized states.

\begin{figure}[tb]
\begin{center}
\includegraphics[width=0.45\textwidth]{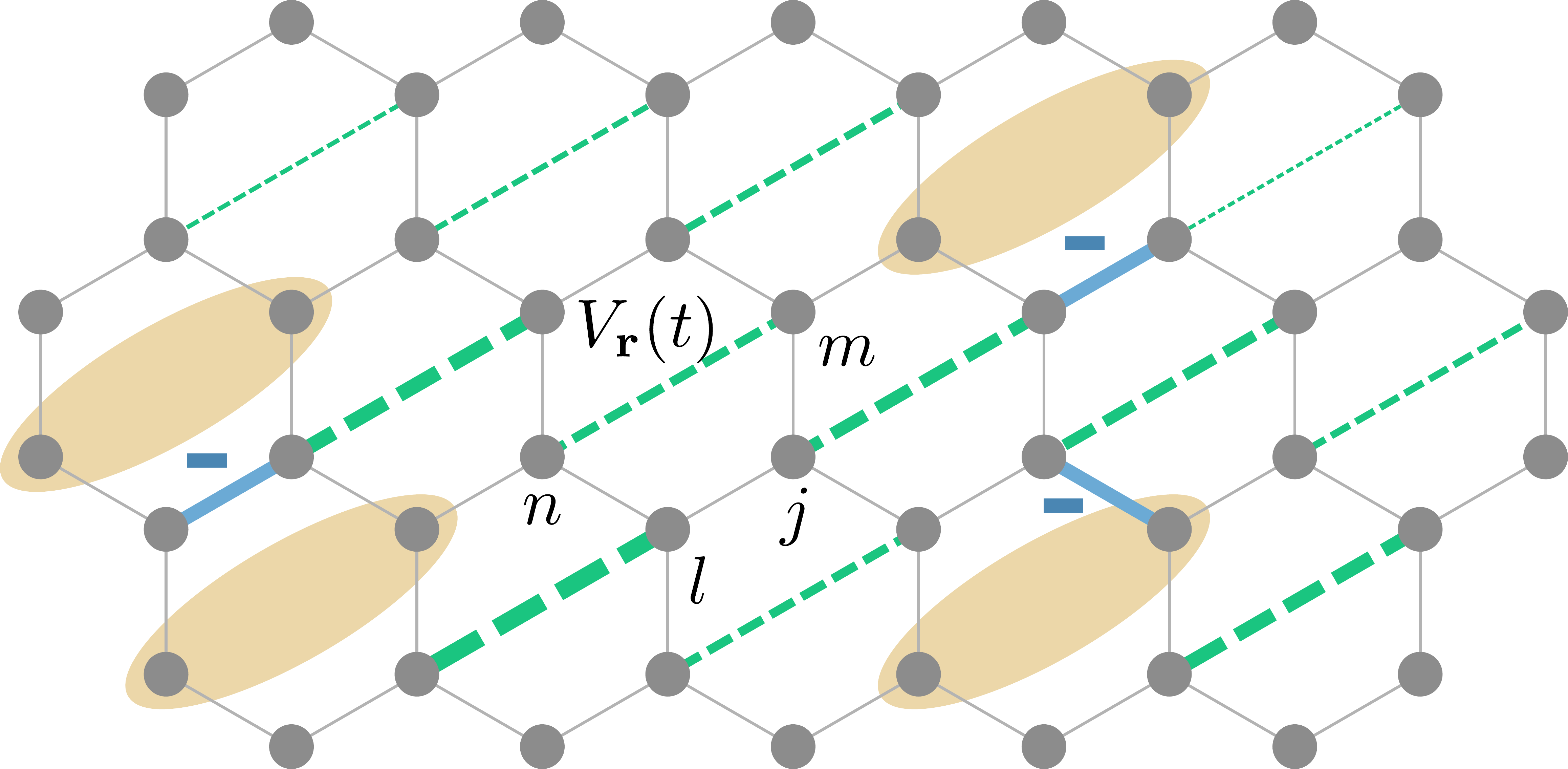}
\caption{The disorder term $V_{\bf r}(t)$ acts only on the localized orbitals of each plaquette and does not lead to delocalization. States in the flat-bands repel in quasi-energy depending on the magnitude of $V_{\bf r}(t)$ (here denoted by the width of the diagonal lines), but remain localized on their respective plaquettes. Floquet-Majorana states with energies $\varepsilon=0,\pi$ formed at vortices (yellow ellipses) also remain unaffected.}
\label{fig:kitaev_v_disorder}
\end{center}
\end{figure}

As a first step, we show that the system remains fully localized in the case of a special type of time-dependent disorder which keeps the model exactly solvable. Starting from $J_u =0$ and $J_s T/4 =\pi/2$, we add a disorder potential that changes the quasi-energies of the clean system, but not the Floquet eigenstates (see Fig.~\ref{fig:kitaev_v_disorder}). The disorder term is given by
\begin{equation}
\label{eq:four_spin}
H_{\mathrm{dis}} = \sum_{n} V_{n}(t)\sigma^{x}_{n}\sigma^{z}_{l}\sigma^{x}_{j}\sigma^{z}_{m}.
\end{equation}
Here, $n$ labels a $B$ sublattice site, and the positions of the sites $l(n)$, $j(n)$, $m(n)$ relative to $n$ are as shown in Fig.~\ref{fig:kitaev_v_disorder}.

In terms of the fermionic representation, $H_{\mathrm{dis}}$ is written as \cite{Po2017}
\begin{equation}
\label{eq:diag_hop}
H_{\mathrm{dis}} = \sum_{n} i V_{n}(t)  c_n c_{m} \, \hat{u}_{nl} \hat{u}_{lj} \hat{u}_{jm}.
\end{equation}
In a fixed gauge, the disorder term acts as a diagonal hopping from a $B$ sublattice site to the $A$ site across the hexagon, with amplitude $V_n(t)$.

To keep the solvability of the model, the time-dependence is chosen such that the disorder acts in a stroboscopic fashion: $V_{n}(t)=0$ for $0<t<3T/4$, and $V_{n}(t) = v_n $ for $3T/4<t<T$.
Here, $v_n$ are independent random numbers drawn from the uniform distribution $\left[-\delta V,\;\delta V\right]$, with $\delta V$ the disorder strength. In the fourth part of the period, the evolution of the eigenmodes localized on the sites $m$, $n$ is
\begin{equation}
\psi_\pm \rightarrow e^{\pm i v_n T / 4} \psi_\pm,
\label{eq:evolution_dis}
\end{equation}
where $\psi_\pm$ are the even and odd combinations of the Majorana operators on the two sites $n$, $m$ across the plaquette, as in Eq.~(\ref{eq:pm}).

In the case of resonant driving, $J_u =0$ and $J_s T/4 =\pi/2$, the Floquet eigenstates are not changed in the presence of this type of disorder. The disorder only changes the quasi-energies of states localized in plaquettes with no $Z_2$ flux, which become $\varepsilon^{\pm}_n = \pm \frac{\pi}{2T} \pm v_n$. The quasi-energies of the pairs of Majorana modes localized on plaquettes with a $Z_2$ flux are \textit{not} affected by the disorder; their quasi-energies are pinned to $0$, $\pi/T$ by particle-hole symmetry. This can be verified by applying the evolution in the first three parts of the period, Eq.~(\ref{eq:pm_vortex}), followed by the evolution due to the disorder, Eq.~(\ref{eq:evolution_dis}), and diagonalizing the resulting evolution operator.

Thus, for resonant driving all the states in the spectrum (in \textit{all} flux sectors) are localized. However, states with a finite density of vortices have a large degeneracy centered at quasi-energies $\varepsilon = 0, \pi/T$, due to the vortex core states (that are not split even in the presence of disorder). Once we deviate from the resonance condition, these states start hybridizing, and we may expect them to delocalize.

\begin{figure}[tb]
 \includegraphics[width=\columnwidth]{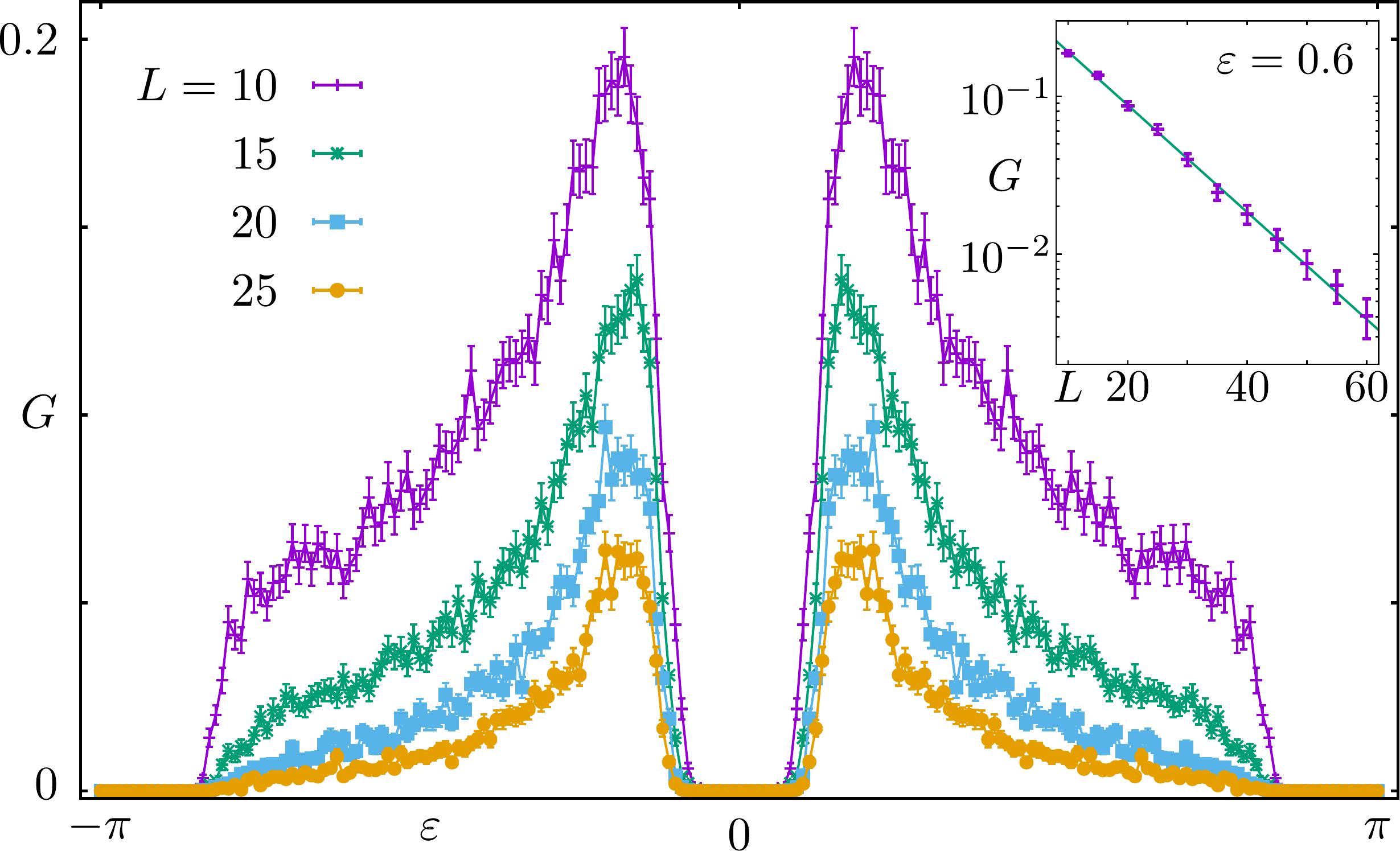}
 \caption{The average two-terminal conductance of a cylindrical system of $L\times L$ unit cells, with periodic boundary conditions along $\vec{a}_x$, is plotted as a function of quasienergy. We use $J_sT/4=1.45$, $J_uT/4=0.1$, and a disorder strength $\delta VT/4=1.0$. The conductance is peaked close to $\varepsilon=0.6$, but decreases with increasing system size at all values of quasienergy, indicating that bulk states are localized. The inset shows a log-linear plot of the average conductance versus system size for a fixed value $\varepsilon=0.6$. The solid line is a fit to an exponential decay $a\exp(-b L)$, with $a\simeq0.42$ and $b\simeq0.08$. Each point is obtained by averaging over 1000 independent realizations of disorder.\label{fig:cond_bulk}}
\end{figure}

In the clean system, moving away from the point of resonant driving while staying in the anomalous phase causes the bulk and Majorana bands to disperse, as states on different plaquettes start to overlap. Upon adding disorder, the bulk bands, exhibiting zero Chern numbers, can be completely localized.
To demonstrate this, we analyze the localization behavior of the bulk states by computing the ``two-terminal conductance'' of the system in a cylindrical geometry (see Appendix \ref{app:scattering_matrix} for details of the transport calculation). The latter is obtained by treating the Majorana modes as if they were complex fermions. The results of Fig.~\ref{fig:cond_bulk} show that the conductance of the bulk states decreases exponentially with system size at all values of quasienergy, signaling that the bulk bands are indeed localized.

\begin{figure}[tb]
\begin{center}
\includegraphics[width=0.35\textwidth]{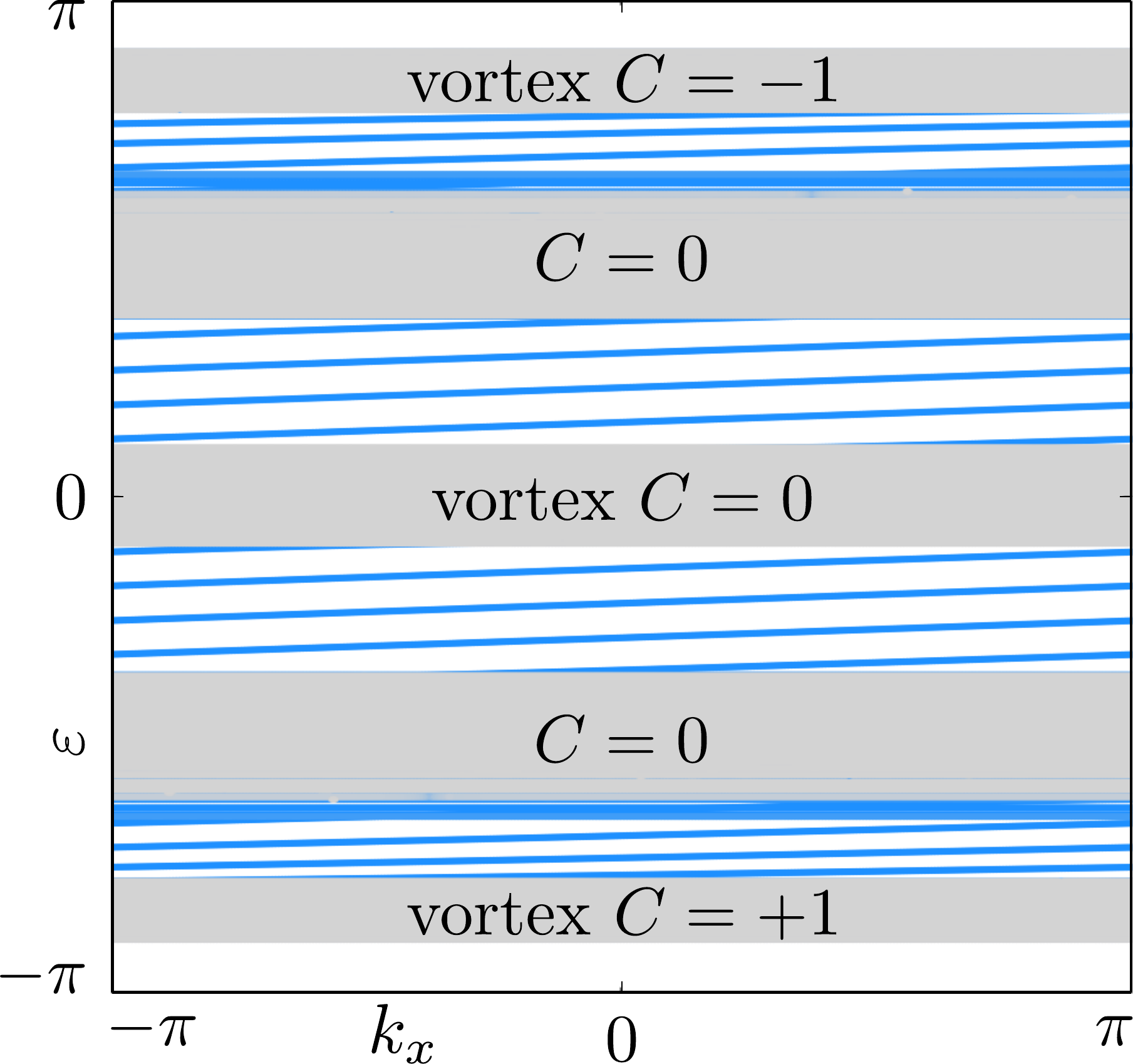}\\
\caption{Quasi-energy spectrum of a $24\times40$ super-cell, which is repeated infinitely many times in the $\vec{a}_x$ direction, such that $k_x$ is a good quantum number. We use a high density $\rho_{v}=0.5$ of randomly distributed vortices, obtained by randomly flipping the sign of each bond with a probability $1/2$, setting $J_{u} T/4=-0.12$ and $J_{s}T/4=\pi/2$. Bulk states are shown in gray, and edge modes in blue. Only states belonging to one edge are plotted.}
\label{fig:lsr_bulk}
\end{center}
\end{figure}

For the bands formed out of Majorana modes, which we call ``vortex bands'' in the following, we find that $0$- and $\pi$-modes show a qualitatively different behavior, depending on the strength and sign of the coupling strengths $J_s$ and $J_u$. We introduce a finite density of flux excitations by randomly flipping the sign of each bond with a probability $\rho_v$. For $J_s T/4=\pi/2$, $J_u T/4=-0.12$, and $\rho_v=0.5$ the 0-modes form a continuous band centered around $\varepsilon=0$, while $\pi$-modes form two symmetric bands, such that a gap develops at $\varepsilon=\pi$ (see Fig.~\ref{fig:lsr_bulk}). Replacing $J_u \to -J_u$, or alternatively $J_sT/4 \to \pi - J_sT/4$, reverses the behavior of Floquet-Majorana states: the system remains gapless at $\varepsilon=\pi$ but shows a gap for $\varepsilon=0$. We remark that these are not exact symmetries of the model for generic values of the hopping strengths.

\begin{figure}[tb]
\begin{center}
\includegraphics[width=0.35\textwidth]{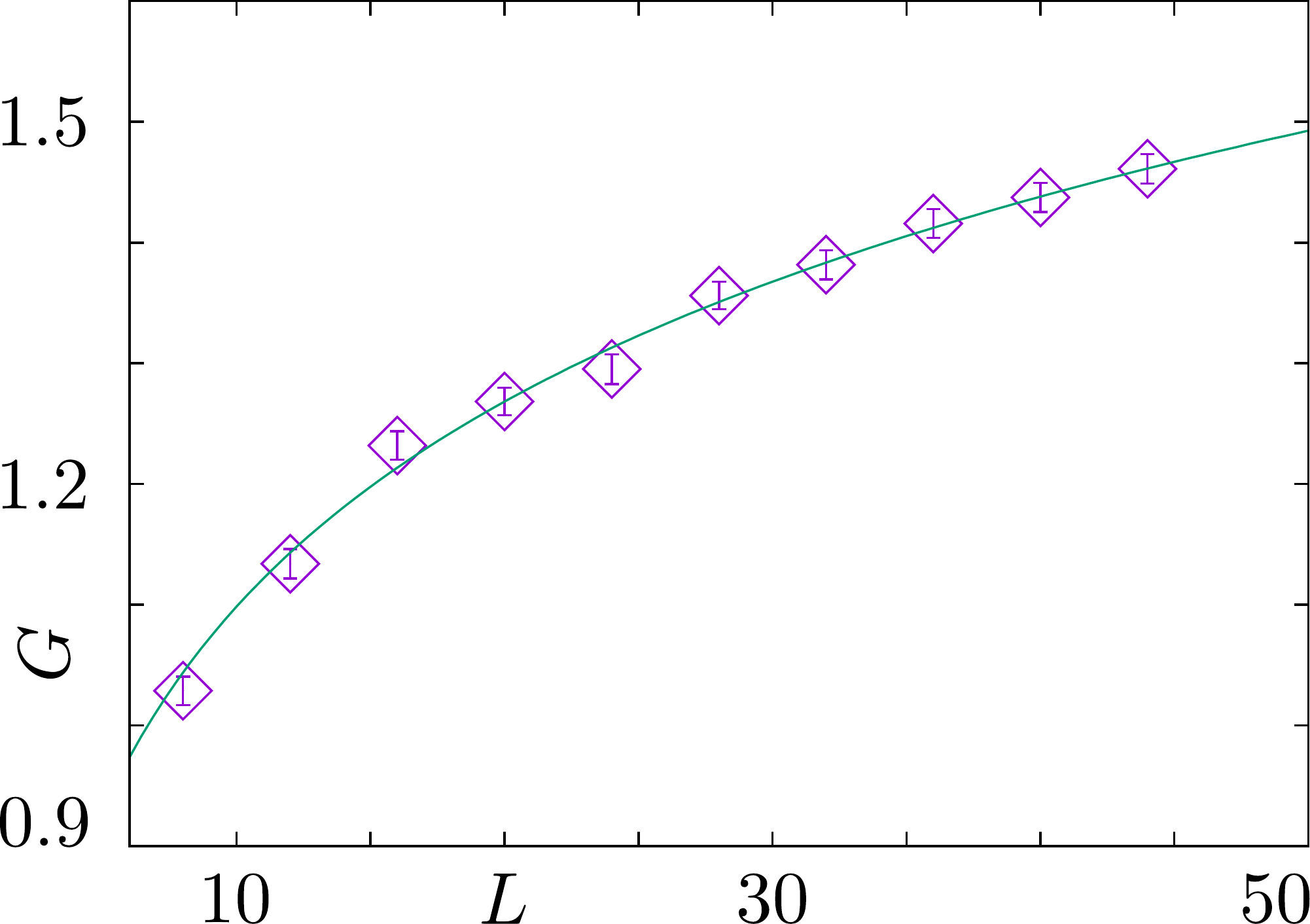}
\caption{Average conductance as a function of system size at $\varepsilon=0$, $J_s T/4=\pi/2$, and $J_u T/4=-0.12$, for an $L\times L$ lattice with periodic boundary conditions in the $\vec{a}_x$-direction. Each point is obtained by averaging over 1000 independent vortex configurations at fixed density $\rho_v=0.5$. The conductance shows a positive scaling with system size, consistent with a thermal metal phase in which $G \propto \log L$ (solid line).}
\label{Fig:ConductanceThermalMetal}
\end{center}
\end{figure}

Both types of vortex band are delocalized in the presence of disorder. The gapped $\pi$-bands of Fig.~\ref{fig:lsr_bulk} carry a non-zero Chern number $C=\pm 1$, as can be directly checked by counting the number of chiral edge states at $\varepsilon=\pi$. Therefore, they are delocalized.\cite{Thouless1984, Thonhauser2006} In contrast, the vortex band around $\varepsilon=0$ is topologically trivial, but shows the features of a delocalized, thermal metal phase. The latter has been shown to appear also in static two-dimensional systems subject to particle-hole symmetry,\cite{Evers2008} irrespective of whether time-reversal symmetry is broken or not,\cite{Medvedyeva2010, Fulga2012a} and is characterized by a logarithmic growth of conductance with system size. We show this behavior in Fig.~\ref{Fig:ConductanceThermalMetal}, where the two-terminal conductance at $\varepsilon=0$ is obtained by averaging over independent vortex configurations at a fixed density of $\rho_v=0.5$ vortices per plaquette. We note here, however, that the existence of the 2D metal is disputable, as discussed in Refs.~\onlinecite{Ziegler2010, Sinner2014, Sinner2016}, which argued that the logarithmic scaling of conductance may be due to finite-size effects. This hypothesis is beyond our ability to test, given the system sizes we can numerically access.

We conclude that at the exactly solvable resonant-driving point of the anomalous topological phase, the Floquet-Kitaev model is fully localized in all vortex sectors. 
However, in the presence of perturbations away from the resonant driving, such as $J_u\neq 0$, the numerical results summarized in Figs.~\ref{fig:lsr_bulk} and \ref{Fig:ConductanceThermalMetal} indicate that in sectors with a non-zero density of $Z_2$ flux excitations, the many-body spectrum contains delocalized states. These states lead either to mobility edges generated by vortex bands with nonzero Chern numbers, or to the formation of a thermal metal phase. Below, we argue that this behavior is {\it generic} in sectors with a finite density of fluxes.

\subsection{Delocalization at finite vortex density}
\label{sec:vortexdens}
We now present a general argument showing that sectors of the many-body Hilbert space with a finite density of $Z_2$ fluxes contain delocalized bulk states. These delocalized bulk states are part of the bands formed by the coupling between the $0$ and $\pi$ Majorana modes bound to the fluxes. Thus, the behavior found numerically in Sec.~\ref{sec:dis_numerics} is generic.

\begin{figure}[t]
\begin{center}
\includegraphics[width=0.3\textwidth]{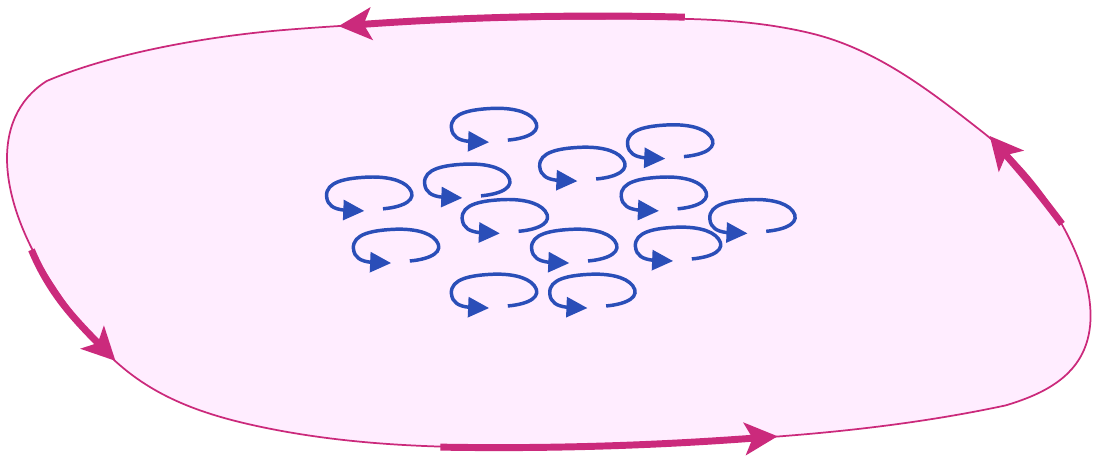}
\caption{Setup discussed in Sec.~\ref{sec:vortexdens}, demonstrating the presence of delocalized states in a region with a finite density of unpaired fluxes. The middle region of a system in the anomalous phase is occupied by a finite density of fluxes, while the outer region in flux free. Depending on microscopic details, the inner region hosts either bands with a non-zero Chern number (in which case a chiral edge state appears at the boundary between the inner and outer regions), or a thermal metal phase.}
\label{Fig:fluxes}
\end{center}
\end{figure}

To see this, we consider the setup
depicted in Fig.~\ref{Fig:fluxes},
where a region near the middle of the system is occupied by a finite density of fluxes. As before, we assume that the fluxes are static, i.e. the flux through every plaquette commutes with the Hamiltonian. The surrounding region is flux-free. The flux positions in the central region are randomly chosen from a certain probability distribution, assumed to satisfy the following conditions: 1) the distribution is translationally invariant in the interior of the central region, as well as along its boundary; 2) the probability to find a flux in a certain plaquette depends only on the presence or absence of other fluxes within a finite circle around that plaquette; 3) for any configuration of fluxes with a non-zero probability, it is possible to find at least one flux whose removal produces a new configuration with a non-vanishing probability. The last condition means that the fluxes do not appear only in ``bound pairs,'' i.e., there is a finite density of {\it unpaired} fluxes\footnote{An example for a distribution that violates this condition is a distribution with a finite density of non-overlapping pairs of fluxes, where the members of every pair occupy nearest-neighbor plaquettes. Then, removing a single flux produces a configuration whose probability is precisely zero. If all the fluxes are paired, their Majorana states {\it can} all be localized. This is since each pair of Majorana modes form a single complex fermion state with a finite on-site energy, and the problem reduces to the usual Anderson model for localization.}.

We now show that in every flux configuration drawn from such a distribution, not all the states in the central region are localized. This follows from considering another flux configuration in which one of the fluxes in the interior of the central region has been removed\footnote{Since the total number of fluxes in the system is always even, this means that a flux must appear at the boundary between the outer region and vacuum.}. Suppose that the old configuration has an even number of fluxes. Since every flux carries a zero mode (a mode at quasienergy $\varepsilon=0$), the new configuration must have one zero mode in the central region (a second zero mode appears in the physical boundary of the system, which we assume to be far away). This zero mode can either be delocalized over the entire central region, delocalized around the boundary between the central region and the outer, flux-free region, or localized around the plaquette of the removed flux. In the latter case, since the distribution of flux configurations is translationally invariant, there must be a {\it finite density} of zero modes in the bulk; such zero modes cannot generically be localized, since every pair of zero modes is precisely in resonance, and any infinitesimal coupling hybridizes them strongly. Therefore, the zero mode generated cannot be localized around the plaquette from which the vortex was removed.  Similar considerations hold for the original configuration if it has an odd number of fluxes.

This implies that the zero mode is either delocalized over the entire central region or around its boundary; since removing a single flux cannot create a new delocalized state, we conclude that either the bulk or the boundary must contain delocalized states essentially for all flux configurations. A delocalized bulk state near zero quasi-energy corresponds to the thermal metal phase. A delocalized state around the boundary implies the existence of a Majorana band with a non-zero Chern number away from zero quasi-energy in the central region; otherwise, there could be no stable delocalized edge state between the interior and the exterior regions (where the Chern numbers of all the bulk states are zero). Since the Majorana band has non-zero Chern number, it must include delocalized bulk states. Thus, in either case, the bands formed by hybridizing the Majorana modes bound to the fluxes must contain delocalized bulk states.

We remark that the argument above holds also for the Majorana bands in the vicinity of quasi-energy $\varepsilon=\pi$. Hence, these bands must also contain delocalized states. Note also that the same argument applies in the case of a static $p_x + ip_y$ topological superconductor, as well. In such a superconductor, vortex cores carry Majorana zero modes; in the mixed state of such a superconductor, the Majorana modes form bands that must contain delocalized states, consistently with the findings of Refs.~[\onlinecite{Grosfeld2006}, \onlinecite{Kraus2011}, \onlinecite{Laumann2012}].

Hence, we conclude that the anomalous phase in the driven Kitaev model cannot be fully localized in all flux sectors. This implies that, strictly speaking, the system cannot be many-body localized. Upon adding more generic perturbations that do not commute with the number of fluxes on every plaquette, the different flux sectors are no longer conserved. One may then expect that the Floquet many-body eigenstates are all delocalized, and the system heats up to infinite temperature at sufficiently long times. However, a subtlety may arise if the system is initially in a state with a zero density of fluxes. Starting from such a state, a macroscopic number fluxes need to be generated in order to delocalize the system; such a process may take a macroscopically long time.
We leave this intriguing possibility to a future study.

\section{Conclusions}
\label{sec:conc}

Periodically driven systems can host different topological phases, including ones that do not have analogues in static systems.
It is natural to ask what new phases exist in strongly interacting systems, such as spin systems.
In this work, we have investigated the topological and localization properties of an interacting spin system, the Kitaev honeycomb model, in the presence of strong driving.

Owing to the exactly solvable nature of the model, which is mapped to a model of free Majorana fermions, we were able to determine its phase diagram, which shows a number of strong and weak topological phases. Among them, the so-called anomalous topological phase is characterized by chiral edge states coexisting with trivial bulk Floquet bands. It shows a number of features unique to periodically driven systems, such as the existence of pairs of Floquet-Majorana bound states at vortices, whose non-Abelian braiding properties resemble those of ordinary Majorana zero modes.

In the anomalous phase, we have studied a form of disorder which keeps the model integrable.
In the flux-free sector, all bulk states remain strictly localized on the hexagonal plaquettes of the lattice due to their topologically trivial nature. Majorana modes bound to isolated vortices are also unaffected by disorder, forming flat bands at quasi-energies $\varepsilon=0$ and $\varepsilon=\pi$.

In the presence of a finite density of vortices, however,
we have shown that the many-body spectrum necessarily contains delocalized states, as a result of the hybridization of the extensive number of Majorana modes centered around quasi-energy $\varepsilon=0$ or $\varepsilon=\pi$.
These Floquet-Majorana modes can either form a thermal metal phase, 
or alternatively form gapped Chern bands, which exhibit mobility edges due to their topologically non-trivial nature.\footnote{This phenomenon is related to the absence of many-body localization in models with certain types of topological order, such as anyon chains. See Ref.~\onlinecite{Potter2016a}}

The fate of the system in the presence of generic interactions, that spoil the integrability of the model, remains an open question. On one hand, since the many-body spectrum always contains delocalized states, one may naively think that many-body localization is impossible - therefore, once generic interactions are introduced, the system necessarily thermalizes.
On the other hand, thermalization requires the presence of a finite density of vortices. If the system is initialized in a vortex-free sector, it may stay localized for a very long time. This is since the matrix elements of any local operator between the initial state and the continuum of delocalized states vanish; the thermalization process is an infinitely high-order process, and thus it may become infinitely slow in the thermodynamic limit. In that case, the topological properties of the system in sectors with a finite number of fluxes are stable over a finite range of generic interactions. We leave a detailed study of this problem to a future study.

Finally, it is interesting to study the topology and localization behavior of the periodically driven Kitaev model in the case of anisotropic coupling strengths $J_{x,y,z}$. While we have considered couplings of equal magnitude for all bonds, a wider range of topological phases may be reached by applying strain or other types of distortions to the hexagonal lattice.

\acknowledgments

The authors thank Adolfo G. Grushin and Zohar Nussinov for helpful discussions. This work was supported by the European Research Council under the European Union's Seventh Framework Programme (FP7/2007-2013) / ERC Project MUNATOP, ERC synergy UQUAM project, the US-Israel Bi-national Science Foundation, ISF grant 1291/12 and the Minerva Foundation.
N. L. and E. B. acknowledge financial support from the European Research Council (ERC) under the European Union Horizon 2020 Research and Innovation Programme (Grant Agreement No. 639172).   N. L. acknowledges support from the People Programme (Marie Curie Actions) of the European Union’s Seventh Framework Programme (No. FP7/2007–2013) under REA Grant Agreement No. 631696, and from the Israeli Center of Research Excellence (I-CORE) ``Circle of Light.'' I. C. F. acknowledges financial support from the DFG through the W{\"u}rzburg-Dresden Cluster of Excellence on Complexity and Topology in Quantum Matter -- \textit{ct.qmat} (EXC 2147, project-id 39085490). We acknowledge support from the DFG under grant CRC 183 (project A01).

\appendix

\section{Scattering matrix formalism}
\label{app:scattering_matrix}

We characterize the topological phases of the periodically driven Kitaev model by using the scattering matrix formalism. The latter is described in detail in Ref.~\onlinecite{Fulga2016}, so we only briefly summarize it here. We consider finite lattices of $L\times W$ sites indexed by pairs of integers $(n_x, n_y)$ and construct a simplified, fictitious scattering problem by defining absorbing terminals on the sites at the top and bottom boundaries of the hexagonal lattice in Fig.~\ref{fig:lattice_protocol}. The absorbers act after each full period of the time evolution, enabling to associate a scattering matrix $S$ to the Floquet operator ${ F}$ as
\begin{equation}
\label{eq:FloquetS}
S(\varepsilon) = P \left[ 1-e^{i\varepsilon}{ F}(1-P^TP) \right]^{-1}e^{i\varepsilon}{ F}P^T,
\end{equation}
where $P$ is the projection operator onto the terminals, and the superscript $T$ denotes transposition.

Owing to the properties of the projection operator and the unitarity of ${ F}$, Eq.~\eqref{eq:FloquetS} defines a unitary, quasi-energy dependent scattering matrix, which for two terminals takes the form
\begin{equation}\label{eq:smatrix}
S = \begin{pmatrix}
r & t \\
t' & r'
\end{pmatrix},
\end{equation}
where $r^{(\prime)}$ and $t^{(\prime)}$ are amplitudes for reflection and transmission between the absorbers, respectively.

The scattering matrix can be used to determine localization properties, obtained by computing the transmission through the system, $G={\rm Tr}\,t^\dag t$, where ${\rm Tr}$ is the trace. Furthermore, it allows to compute the topological invariants of a phase, which take the same form as in static systems,\cite{Braeunlich2009, Fulga2011, Fulga2012, Tarasinski2014} even in the anomalous phase in which all bulk Floquet bands are topologically trivial. The strong invariant is obtained by imposing twisted boundary conditions in the direction perpendicular to the terminals, connecting the left and right boundaries of Fig.~\ref{fig:lattice_protocol} as $|L,n_y\rangle = e^{i\phi}|0, n_y\rangle$, with $\phi$ the twist angle. The reflection sub-block $r$ of the scattering matrix in Eq.~\eqref{eq:smatrix} is now a function of both $\phi$ and quasi-energy $\varepsilon$, enabling to write the strong topological index as the winding number of $\det r$:
\begin{equation}\label{eq:rchern}
{\cal W}_\varepsilon = \frac{1}{2\pi i}\int_0^{2\pi} d\phi\,\frac{d}{d\phi}\log\,\det\,r(\varepsilon, \phi).
\end{equation}

In addition to strong topological phases, the periodically driven Kitaev model also shows weak phases, protected by a combination of translation and particle-hole symmetry Eq.~\eqref{eq:ph_floquet_momentum_space}. Given Eq.~\eqref{eq:FloquetS}, particle-hole symmetry constrains the scattering matrix as
\begin{equation}\label{eq:ph_sym_S}
S(\varepsilon) = S^*(-\varepsilon)
\end{equation}
for both periodic ($\phi=0$) and anti-periodic ($\phi=\pi$) boundary conditions, meaning that the scattering matrix is real at the particle-hole symmetric quasi-energies $\varepsilon=0,\pi$. This enables us to write weak topological indices as
\begin{equation}\label{eq:phs_weak_inv}
\begin{split}
\nu_{x,0,\varepsilon=0} &= {\rm sign}\,\det r(\varepsilon=0,\phi=0), \\
\nu_{x,\pi,\varepsilon=0} &= {\rm sign}\,\det r(\varepsilon=0, \phi=\pi),
\end{split}
\end{equation}
where $\phi=0$ and $\phi=\pi$ denote periodic and anti-periodic boundary conditions, respectively, and the subscript $x$ refers to the direction in which they are imposed. In a similar fashion, one can define weak invariants for the other particle-hole symmetric quasi-energy, $\varepsilon=\pi$, as well as indices associated to the edges parallel to $a_y$: $\nu_{y,0/\pi}$. The latter are obtained by defining absorbing terminals on the left and right boundaries of Fig.~\ref{fig:lattice_protocol} and imposing (anti-)periodic boundary conditions along $a_y$.

The strong invariant Eq.~\eqref{eq:rchern} is defined for all quasi-energies at which there is a bulk mobility gap, such that $\det r(\phi)\neq 0$ for all $\phi$, and counts the net number of chiral edge states. In contrast, weak invariants Eqs.~\eqref{eq:phs_weak_inv} can only be defined at $\varepsilon=0,\pi$, due to the constraint Eq.~\eqref{eq:ph_sym_S}, and count the parity of edge modes present at $k_i=0$ and $k_i=\pi$, with $i=x,y$. At a fixed quasi-energy, weak invariants in the same direction are not independent, being related by the parity of the strong index: $\nu_{i,0, \varepsilon}\nu_{i,\pi,\varepsilon}=(-1)^{{\cal W}_\varepsilon}$, with $\varepsilon=0,\pi$ and $i=x,y$.\cite{Ran2010} As such, there are in total six invariants characterizing a phase, $({\cal W}_{\varepsilon=0}, \nu_{x,\pi,\varepsilon=0}, \nu_{y,\pi,\varepsilon=0}, {\cal W}_{\varepsilon=\pi}, \nu_{x,\pi,\varepsilon=\pi}, \nu_{y,\pi,\varepsilon=\pi})$, which are in general independent of each other,\cite{Diez2015} and a change in any one of them is accompanied by a closing of the bulk mobility gap at the associated quasi-energy. However, in our driving protocol both the uniform hopping term $J_u$ as well as the stroboscopic term $J_s$ are chosen to be isotropic, leading to a model in which weak indices along the $x$ and $y$ directions are equal. We are therefore left with four independent invariants, which are used to label the topological phases of Fig.~\ref{fig:phase_diag}.

\section{Equality of $x$ and $y$ weak invariants}
\label{app:ribbon}

\begin{figure*}[tb]
 \includegraphics[width=1.75\columnwidth]{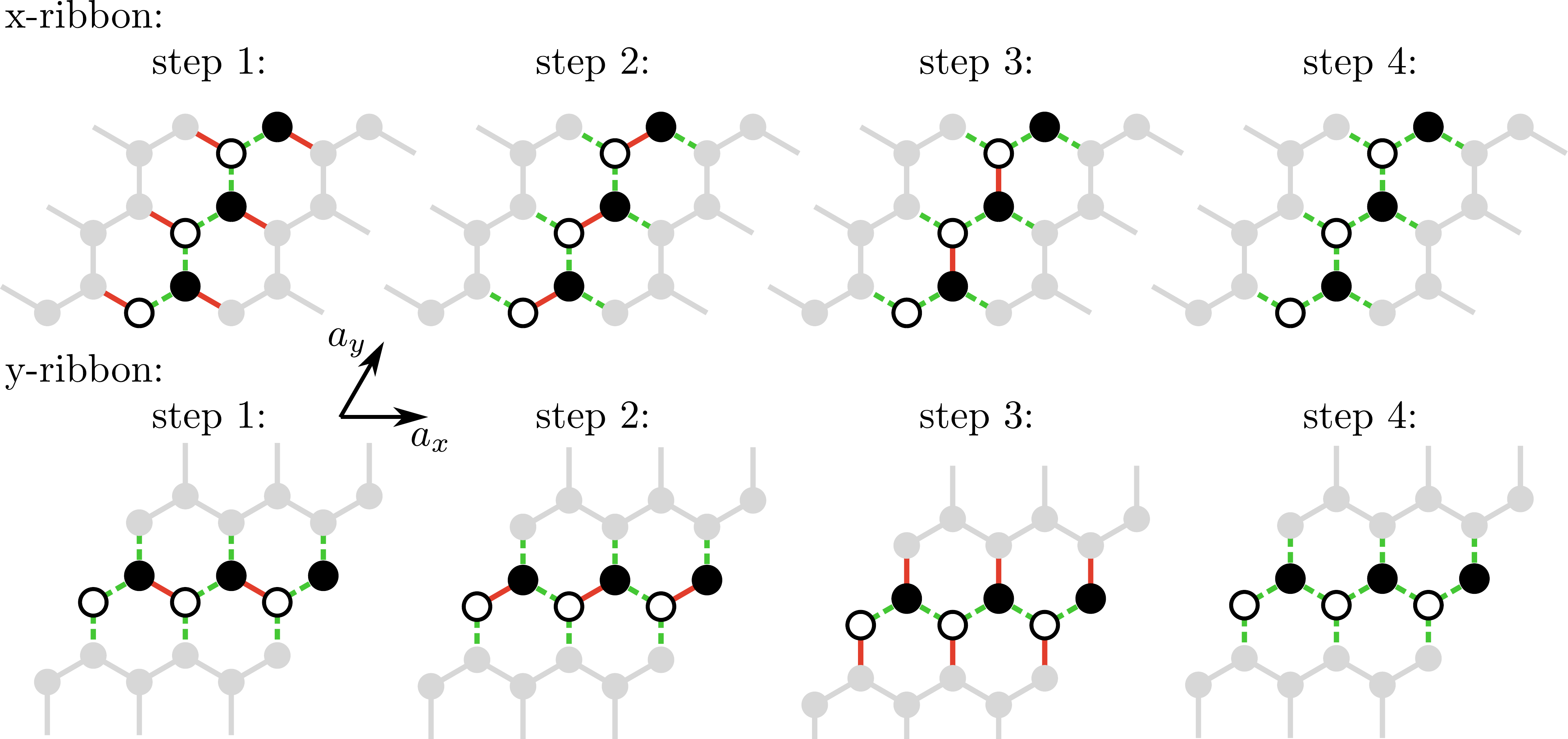}
 \caption{Schematic representation of the Hamiltonians $H_{x/y,i}(k)$ of the Floquet-Kitaev model during the four steps of the driving protocol, for a ribbon infinite along $a_x$ (top row) or $a_y$ (bottom row). For both geometries, the unit cell of the ribbon consists of 6 sites, shown as white and black circles ($A$ and $B$ sublattices, respectively). Adjacent unit cells are shown in light gray. During each step of the time-evolution, nearest neighbor hoppings may take the value $J_u$ (green dashed lines) or $J_u+J_s$ (red solid lines). From the structure of the hoppings it follows that the Hamiltonians of the x-ribbon are identical to those of the y-ribbon, but differ in their order. Steps 2 and 4 are identical, $H_{x,2}(k)=H_{y,2}(k)$ and $H_{x,4}(k)=H_{y,4}(k)$, whereas steps 1 and 3 are interchanged, $H_{x,1}(k)=H_{y,3}(k)$ and $H_{x,3}(k)=H_{y,1}(k)$. \label{fig:strips}}
\end{figure*}

Here we prove that, in any phase of the Floquet-Kitaev model, the weak invariants $\nu_{x,k,\varepsilon}$ and $\nu_{y,k,\varepsilon}$ are equal. We consider the system in two different ribbon geometries, one with zig-zag edges parallel to the $a_x$ direction (which we refer to as the ``x-ribbon''), and one with zig-zag edges parallel to $a_y$ (the ``y-ribbon''). Bandstructures obtained from an x-ribbon are shown both in Fig.~\ref{fig:phase_diag} and in Fig.~\ref{fig:strobo_periodicity}. We will show that, for any value of  $J_s$ and $J_u$, the y-ribbon bandstructures are the same as those of the x-ribbon up to a mirror operation. Therefore, in any gapped phase the edge mode structure is the same in both geometries, which immediately implies equal weak invariants, $\nu_{x,k,\varepsilon}=\nu_{y,k,\varepsilon}$.

Unlike the Floquet operator of the bulk system, which is a $2\times2$ matrix obtained from Eq.~\eqref{eq:floquet_decompose}, in a ribbon geometry both the Hamiltonian and the Floquet operator have a size $N\times N$, where $N$ is the number of sites in the finite direction of the ribbon. The Floquet operator of the x-ribbon reads
\begin{equation}\label{eq:xribbon}
 F_x(k)=e^{-i\frac{T}{4}H_{x,4}}e^{-i\frac{T}{4}H_{x,3}}e^{-i\frac{T}{4}H_{x,2}}e^{-i\frac{T}{4}H_{x,1}},
\end{equation}
where $H_{x,i}(k)$ are the $N\times N$ ribbon Hamiltonians during the four steps of the driving protocol, and $k$ is the momentum along the infinite direction of the ribbon. Similarly, $F_y(k)$ is obtained as a product of exponentials of $H_{y,i}(k)$. Our goal in the following is to show that the spectra of $F_x(k)$ and $F_y(-k)$ are identical.

Fig.~\ref{fig:strips} shows the four steps of the driving sequence of x- and y-ribbons consisting of $N=6$ sites along the finite direction. We observe that, for any given step, the Hamiltonian of the x-ribbon is identical to one of the Hamiltonians characterizing the time-evolution of the y-ribbon. For instance, during the first step of the x-ribbon, hoppings connecting sites inside the same $6\times6$ unit cell take the value $J_u$ (dashed green lines), whereas hoppings connecting neighboring unit cells are equal to $J_u+J_s$ (red solid lines). The same structure of hoppings can be found during the third driving step of the y-ribbon, such that $H_{x,1}(k)=H_{y,3}(k)$. Another way to see this is to note that the two lattices depicting $H_{x,1}(k)$ and $H_{y,3}(k)$ can be mapped onto each other by a $60^\circ$ rotation followed by a mirror operation, such that the left edge of the y-ribbon becomes the bottom edge of the x-ribbon.
For the narrow ribbons shown in Fig.~\ref{fig:strips}, the two Hamiltonians can be written down explicitly, enabling us to verify their equivalence:
\begin{equation}
H_{x,1}(k) = H_{y,3}(k)=
 \begin{pmatrix}
    0 & t_1^* &     0 & 0 & 0 & 0 \\
  t_1 &     0 & t_2^* & 0 & 0 & 0 \\
    0 &   t_2 &     0 & t_1^* & 0 & 0 \\
    0 & 0     &   t_1 & 0 & t_2^* & 0 \\
    0 & 0     &     0 & t_2 & 0 & t_1^* \\
    0 & 0     &     0 & 0 & t_1 & 0 \\
 \end{pmatrix},
\end{equation}
where $t_1 = iJ_u + i(J_u+J_s)e^{-ik}$ and $t_2 = -iJ_u$.

It follows that the Floquet operators $F_x(k)$ and $F_y(k)$ are constructed from the same Hamiltonians, but contain a different order of steps with respect to each other. Specifically, as can be seen from Fig.~\ref{fig:strips}, the only difference is that steps 1 and 3 are interchanged, such that
\begin{equation}\label{eq:hequiv}
 \begin{split}
  H_{x,1}(k) & = H_{y,3}(k), \\
  H_{x,2}(k) & = H_{y,2}(k), \\
  H_{x,3}(k) & = H_{y,1}(k), \\
  H_{x,4}(k) & = H_{y,4}(k). \\
 \end{split}
\end{equation}
Furthermore, since the model consists of a bipartite lattice of Majorana modes, during each step of the driving protocol the ribbon obeys both a particle-hole symmetry,
\begin{equation}\label{eq:phs_ribbon}
 H^\pd_{d,i}(k) = -H^*_{d,i}(-k) = -H^T_{d,i}(-k),
\end{equation}
with $d=x,y$ and the superscript $T$ denoting transposition, as well as a sublattice symmetry,
\begin{equation}
 H_{d,i}(k) = - SH_{d,i}(k)S,
\end{equation}
with $S={\rm diag}(1, -1, 1, -1, \ldots, 1, -1)$. Using the above two symmetries as well as Eqs.~\eqref{eq:hequiv}, it can be shown that $F_x(k)$ and $F_y(-k)$ have identical eigenphases, since
\begin{equation}\label{eq:fl_equiv}
F^\pd_x(k) = U_{xy}(k) F_y^T(-k) U_{xy}^\dag(k),
\end{equation}
where $U_{xy}(k) = \exp\left[-i H_{x,4}(k)T/4\right] S$.

Notice that Eq.~\eqref{eq:fl_equiv} implies that the bandstructures of the x- and y-ribbons are mirrored with respect to each other, since the transformation between them involves changing the sign of momentum, $k\to-k$. As such, in phases with non-zero strong invariants, the velocity of chiral modes will be flipped when mapping an x-ribbon to a y-ribbon. This is consistent with their uni-directional nature however, since the left boundary of the y-ribbon is transformed into the bottom boundary of the x-ribbon. For example, a clockwise propagating chiral mode will propagate in the $-a_x$ direction on the bottom boundary, but in the $+a_y$ direction on the left boundary. Flipping the sign of momentum however does not change the weak invariants, which count the parity of the number of edge modes present at $k=0,\pi$.

\section{Quasi-periodicity of the phase diagram}
\label{apdx:periodic_phase_diag}

\begin{figure}[tb]
\begin{center}
\includegraphics[width=0.45\textwidth]{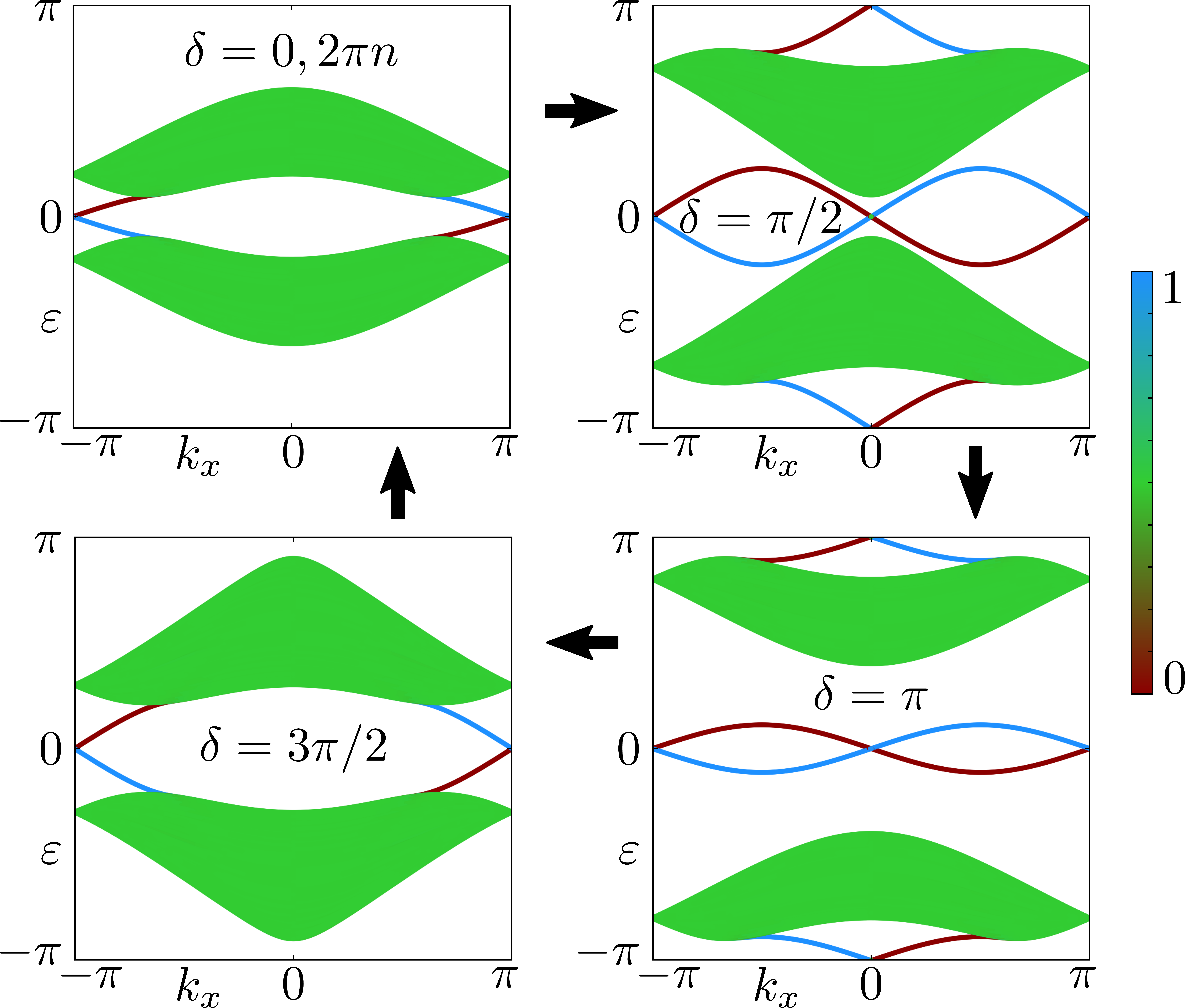}
\caption{ For $J_{u}=0$ phase diagram is $2\pi$-periodic in $J_{s}T/4$. For $J_{s}T/4=\pi/5+\delta$ spectrum returns to itself as $\delta=2\pi n$. The banstructures are computed in an infinite ribbon geometry (infinite along $\vec{a}_x$, width $W=40$) and the color scale indicates the wavefunction amplitude on the bottom 20 lattice sites.}
\label{fig:strobo_periodicity}
\end{center}
\end{figure}

The stroboscopic nature of the driving protocol leads to a quasi-periodic structure of the phase diagram for $J_{u}\ll J_{s}$, as can be seen in Fig.~\ref{fig:phase_diag}.
This periodicity is exact for $J_{u}=0$ and can be understood at the resonant driving points.
If $J_{s}T/4=\pi/2$, during each segment of the driving a Majorana is transfered with unit probability to a neighboring site. After two driving periods it returns to its origin, resulting in the formation of two flat bands at quasi-energies $-\pi/2$ and $\pi/2$. For $J_{s}T/4=\pi$ Majoranas are transfered with unit probability to the next site and then back, such that they return to the original position at the end of the period. The quasi-energy of the flat bands of Floquet eigenmodes is $\varepsilon=\pi$. Each subsequent increase of $J_sT/4$ by $\pi/2$ corresponds to another full transfer of a Majorana operator between the two neighboring sites. As such, the Floquet operator is $2\pi$-periodic:
\begin{equation}
{\mathcal F} \left({J_{s}T/4}\right)={\mathcal F} \left({J_{s}T/4+2\pi n}\right).
\end{equation}
This procedure generally allows to choose an arbitrarily slow driving protocol, while still being able to observe the full range of topological phases (see Fig.~\ref{fig:strobo_periodicity}).

\section{Non-abelian statistics of Majorana pairs}
\label{app:braiding}

In time-independent systems Majorana modes exhibit non-Abelian statistics, such that interchanging two or more zero modes amounts to performing a unitary transformation on the degenerate subspace spanned by their fermionic parities. One of the key ingredients enabling this possibility is the existence of an energy gap separating the many-body ground state manifold from that of excited states. In the presence of such a gap, the adiabatic theorem guarantees that a sufficiently slow braiding process will not generate unwanted excitations and the system will remain in its ground state at all times.

In a periodically driven system, one cannot speak of a ground state since energy is not conserved. However, it is still possible to consider the braiding properties of Floquet-Majorana modes by making use of the Floquet adiabatic theorem (see Appendix \ref{app:adiabatic} for details).
According to the latter, a sufficiently slow braiding process implemented starting from an initial Floquet \textit{reference state} will not generate additional excitations relative to that state. Therefore, at any point during the braiding process the system will remain in the degenerate manifold of that particular reference state.

In the following we will study the non-Abelian statistics of Floquet-Majorana modes occurring in the anomalous phase of the periodically-driven Kitaev model. In particular, we will show that even though a single ${\mathbb Z}_2$ flux traps a pair of Majoranas, one at $\varepsilon=0$ and one at $\varepsilon=\pi$, the two types of zero modes behave as two \textit{independent} sets of anyons: the fermion parity associated with each type of anyon is separately conserved. This is in contrast to static time-reversal invariant systems, in which Kramers pairs of Majorana modes are not generally protected against such parity transfers.\cite{Wolms2016} We stress that throughout this appendix we will only examine the system in the presence of a finite number of vortices, which implies a vanishing flux density in the thermodynamic limit.

To proceed, we consider the driven Kitaev model in its anomalous phase and in the presence of two well-separated ${\mathbb Z}_2$ fluxes, labeled by $j=1,2$, each of which hosts a pair of Majorana modes $\gamma_j^{(0)}$ and $\gamma_j^{(\pi)}$ at quasi-energy $\varepsilon=0$ and $\varepsilon=\pi$, respectively. These vortex states are separated by a quasi-energy gap from bulk modes, which enables to define a Floquet reference state whose degenerate manifold is spanned \textit{only} by the parities of the Majorana modes: a requirement for any protected braiding operation. A necessary condition for such a state is that the bulk Floquet bands must be either fully occupied or fully empty. To see this, note that if any Floquet band were only partially filled, the braiding process could change the occupation numbers of states in that band at vanishing quasi-energy cost, thus producing entanglement between bulk and vortex modes. For concreteness, we will use an initial Floquet state in which the bulk band at negative quasi-energies is fully filled, while the one at positive energies is completely empty.

To implement the vortex interchange, we consider a family of time-periodic Hamiltonians $H(\lambda, t)=H(\lambda, t+T)$ and associated Floquet operators ${ F}(\lambda)$ with $0\leq \lambda \leq 1$ parameterizing the slow braiding process, $\dot\lambda \ll 1/T$. We choose the Floquet operators such that they coincide at the beginning and end of the interchange process, ${ F}(0)={ F}(1)$, as is common for braiding operations also in static systems.

Due to the Floquet adiabatic theorem (see Appendix \ref{app:adiabatic}), transitions outside of the degenerate manifold (which the reference state is a part of) are suppressed as long as the braiding process is slow compared to the quasi-energy gap, $\dot\lambda \ll \Delta\varepsilon$. This means that the occupation numbers of the bulk states cannot change, since the bulk bands are separated from each other and from vortex states by quasi-energy gaps and since they are assumed to be either completely fully or completely empty. Another immediate consequence of the existence of quasi-energy gaps is that neither the parity of $0$-modes nor that of $\pi$-modes can change, and the two sets of Majoranas behave as independent anyons. Therefore, their braiding properties can be deduced in a manner analogous to that of Majorana modes in static systems. After completing the vortex interchange, the operators transform as:
\begin{align}
	\gamma^{(0)}_{1} &\to s_{0,1} \gamma^{(0)}_{2} \, ,	\nonumber \\
	\gamma^{(\pi)}_{1} &\to s_{\pi,1} \gamma^{(\pi)}_{2} \, ,	\nonumber \\
	\gamma^{(0)}_{2} &\to s_{0,2} \gamma^{(0)}_{1} \, ,	\nonumber \\
	\gamma^{(\pi)}_{2} &\to s_{\pi,2} \gamma^{(\pi)}_{1} \, ,
\label{eq: transformation}
\end{align}
where $s_j=\pm 1$ since the Majorana operators are hermitian and therefore cannot pick up complex phases. Furthermore, by performing the gauge transformation $\gamma_2^{(0)} \to s_{0,1} \gamma_2^{(0)}$ and $\gamma_2^{(\pi)} \to s_{\pi,1} \gamma_2^{(\pi)}$ we can always set $s_{0,1}=s_{\pi,1}=1$.

Importantly, if one of the remaining signs was positive, $s_{\alpha,2}=1$, the parity of Majorana modes at quasi-energy $\alpha$ would be violated. To see this, note that if $s_{\alpha,2}=1$ then the fermionic creation operator $d^\dag_\alpha = \gamma_1^{(\alpha)}+i\gamma_2^{(\alpha)}$ transforms as $d^\dag_{\alpha}\to id_{\alpha}$ after the vortex interchange. As such, the parity $p_\alpha = 2d^\dag_{\alpha}d_{\alpha} \to - p_{\alpha}$ is explicitly broken. We conclude that the remaining signs must be negative, $s_{\alpha, 2}=-1$, such that Eq.~\eqref{eq: transformation} describes a braiding process identical to that of static systems, with the Majorana zero and $\pi$-modes behaving as independent anyons.

\section{Floquet adiabatic theorem}
\label{app:adiabatic}

In this section, we briefly review the quantum adiabatic perturbation theory, closely following the discussion of Ref.~\onlinecite{Weinberg2017} and using the same notation. For ease of presentation we begin by discussing the adiabatic theory in the case of static Hamiltonians, before moving to the periodically driven systems discussed in the main text.

Consider a gapped system initialized in its non-degenerate ground state and subjected to a Hamiltonian $H(\lambda)$, where $\lambda$ is a slowly varying parameter. One possible example is to consider a topological superconductor with a fixed fermionic parity and only two well-separated Majorana zero-modes, in which case $\lambda$ would describe the slow interchange of the Majoranas. It is convenient to go to a moving frame with respect to $\lambda$ by defining rotated states $|\widetilde{\psi}\rangle = V^\dag (\lambda) |\psi\rangle$, where the unitary operator $V(\lambda)$ diagonalizes the instantaneous Hamiltonian as $\widetilde{H}(\lambda) = V^\dag(\lambda) H(\lambda) V(\lambda)$. Applying this rotation to the time-dependent Schr{\" o}dinger equation ($\hbar=1$ throughout)
\begin{equation}\label{eq:schrod}
 i \frac{d}{dt} |\psi \rangle = H(\lambda) |\psi\rangle
\end{equation}
leads to
\begin{equation}\label{eq:movingschrod}
 i \frac{d}{dt} |\widetilde{\psi} \rangle = \widetilde{H}_m(\lambda) |\widetilde{\psi}\rangle,
\end{equation}
where the so called moving frame Hamiltonian
\begin{equation}\label{eq:movingH}
 \widetilde{H}_m(\lambda) = \widetilde{H}(\lambda) - {\dot\lambda} \widetilde{\cal A}(\lambda)
\end{equation}
and $\widetilde{\cal A}(\lambda) = i V^\dag(\lambda) \partial_\lambda V(\lambda)$ is known as the adiabatic gauge potential. If there are no level crossings for any value of $\lambda$, then $\widetilde{H}(\lambda)$ cannot lead to transitions between the instantaneous eigenstates, since it is diagonal by definition. The only possible transitions then occur due to the adiabatic gauge potential, and have amplitudes that can be estimated using first order perturbation theory. The transition amplitude between two \textit{instantaneous} eigenstates $|n(\lambda)\rangle$ and $|m(\lambda)\rangle$ reads:\cite{Weinberg2017}
\begin{equation}\label{eq:transamp}
 \alpha_{nm} \propto \frac{\dot\lambda}{E_n - E_m} \langle n(\lambda)|V(\lambda) \widetilde{\cal A} (\lambda)V^\dag(\lambda) |m(\lambda) \rangle,
\end{equation}
where the unitary operator $V(\lambda)$ is responsible for rotating the states to a $\lambda$-independent basis, and $E_{n,m}$ are the energies of the two levels. As expected, level transitions are suppressed if the system is varied slowly with respect to its energy gap, \textit{i.e.} when $\dot\lambda \ll \Delta E$.

The same qualitative result is obtained when applying adiabatic perturbation theory to Floquet systems, although its derivation is slightly more involved. As in the main text, we now consider a family of Hamiltonians $H(\lambda, t)=H(\lambda, t+T)$ which depend periodically on time for a fixed value of $\lambda$, but which also have a second, implicit time dependence due to the fact that $\lambda$ itself changes slowly in time. Again following Ref.~\onlinecite{Weinberg2017}, we begin by considering the system at a fixed value of $\lambda$, and use Floquet's theorem to write the time evolution operator as
\begin{equation}\label{eq:tevol}
 U(t, 0) = P(t) \exp [-i t H_F] P^\dag (0),
\end{equation}
where $P(t)=P(t+T)$ is the time-periodic micromotion operator, which describes the time evolution within a driving period, and $H_F$ is the so called Floquet Hamiltonian. Note that both operators depend on the choice of initial time. However, this choice is just a gauge degree of freedom, so for simplicity we choose an initial time $t_0=0$ which means setting $P(0)=1$ throughout the following.

Going to a moving frame with respect to the micromotion operator, again for fixed $\lambda$, allows to write the Floquet Hamiltonian as
\begin{equation}\label{eq:flham}
 H_F(\lambda) = P^\dag(\lambda, t)H(\lambda, t)P(\lambda, t) - i P^\dag(\lambda, t)\frac{\partial}{\partial t}P(\lambda, t),
\end{equation}
with eigenstates defined as
\begin{equation}\label{eq:flhamevec}
 H_F(\lambda)|n_F(\lambda)\rangle = \varepsilon_n(\lambda) |n_F(\lambda)\rangle.
\end{equation}
Note that for fixed $\lambda$ the Floquet Hamiltonian is a time-independent operator, but one which has an unbounded spectrum, since the quasi-energies are only defined modulo $2\pi$.

As in the static case, we can evaluate the effect of a slowly varying $\lambda$ by considering the moving frame of the Floquet Hamiltonian, \textit{i.e.} the unitary operator $V(\lambda)$ which diagonalizes $\widetilde{H}_F(\lambda) = V^\dag(\lambda)H_F(\lambda)V(\lambda)$ for every $\lambda$. To reach this moving frame however, two separate rotations are needed: the first with respect to the micromotion operator and the second with respect to $\lambda$. By making the substitution
\begin{equation}
 |\widetilde{\psi}\rangle = V^\dag(\lambda)P^\dag(\lambda,t)|\psi\rangle
\end{equation}
in the time-dependent Schr{\"o}dinger equation we obtain
\begin{equation}\label{eq:schrodfl}
 i\frac{d}{dt}|\widetilde{\psi}\rangle = \left( \widetilde{H}_F(\lambda) - \dot\lambda \widetilde{\cal A}_F(\lambda, t) \right) | \widetilde{\psi}\rangle.
\end{equation}
As before, an adiabatic gauge potential $\widetilde{\cal A}_F(\lambda, t)$ is responsible for transitions between instantaneous eigenstates, but in the Floquet setting it takes a more complicated form:
\begin{equation}\label{eq:floqA}
\begin{split}
 \widetilde{\cal A}_F(\lambda, t) & = iV^\dag(\lambda)\frac{\partial}{\partial\lambda}V(\lambda) \\ &+iV^\dag(\lambda)P^\dag(\lambda,t)\left[ \frac{\partial}{\partial\lambda}P(\lambda,t) \right]V(\lambda).
\end{split}
\end{equation}
The first part of Eq.~\eqref{eq:floqA} has the same interpretation as in the static case, and describes level transitions due to the slowly varying $\lambda$. The second one is unique to Floquet systems and describes transitions produced by the fact that, when $\lambda$ changes in time, the micromotion operator is no longer periodic.

Perturbation theory shows that transition amplitudes will be suppressed as the difference between quasi-energies increases. However, the spectrum of the Floquet Hamiltonian is unbounded and contains infinitely many physically equivalent quasi-energy levels. To take all possible transitions between them into account, we Fourier transform the gauge potential as
\begin{equation}\label{eq:Afourier}
 \widetilde{\cal A}_F(\lambda, t) = \sum_{q=-\infty}^{+\infty} e^{2\pi i q t/T} \widetilde{\cal A}_{F}(\lambda, q),
\end{equation}
with $q$ an integer labeling the harmonics. Then the transition amplitudes are given by
\begin{equation}\label{eq:alphaFl}
\begin{split}
 \alpha_{mn} \propto & \sum_{q=-\infty}^{+\infty} e^{2\pi i q t/T} \frac{\dot\lambda}{\varepsilon_n-\varepsilon_m+2\pi q/T}\times \\
 & \langle n_F(\lambda) | V(\lambda)\widetilde{\cal A}_F(\lambda, q) V^\dag(\lambda) | m_F(\lambda) \rangle.
\end{split}
\end{equation}
Therefore, if a Floquet reference state is separated by a quasi-energy gap $\Delta\varepsilon$ from other states and the braiding process is slow compared to this gap, $\dot\lambda\ll\Delta\varepsilon$, then level transitions are suppressed.

\bibliography{flwti}

\end{document}